\documentclass[12pt,a4paper]{article}
\usepackage{amsfonts,latexsym}
\usepackage[dvips]{graphicx}
\usepackage{epsfig}
\usepackage{html}

\font\mybb=msbm10 at 12pt
\def\bb#1{\hbox{\mybb#1}}
\def\Z {\bb{Z}}
\def\R {\bb{R}}

\makeatletter
\@addtoreset{equation}{section}
\makeatother

\def\unit{\hbox to 3.3pt{\hskip1.3pt \vrule height 7pt width .4pt \hskip.7pt
\vrule height 7.85pt width .4pt \kern-2.4pt
\hrulefill \kern-3pt
\raise 4pt\hbox{\char'40}}}

\def\x{\times}

\def\be{\begin{equation}}
\def\ee{\end{equation}}
\def\bea{\begin{eqnarray}}
\def\eea{\end{eqnarray}}

\def \alfa {2 \pi \alpha'}
\def \ti {\tilde}
\newcommand{\w}[1]{\\[0.#1cm]}

\begin{document}
\begin{flushright}
\footnotesize
UG-15/98\\
QMW-PH-98-39 \\
SPIN-98/14 \\
{\bf hep-th/9812224}\\
December, $1998$
\normalsize
\end{flushright}

\begin{center}


\vspace{.2cm}
{\LARGE {\bf Spacetime--Filling Branes and Strings with Sixteen Supercharges
}}


\vspace{1cm}

{\bf E.~Bergshoeff}
\footnote{E-mail address: {\tt E.Bergshoe@phys.rug.nl}},
{\bf E.~Eyras}
\footnote{E-mail address: {\tt E.A.Eyras@phys.rug.nl}},
{\bf R.~Halbersma}
\footnote{E-mail address: {\tt R.Halbersma@phys.rug.nl}},
{\bf J.P.~van der Schaar}
\footnote{E-mail address: {\tt J.P.van.der.schaar@phys.rug.nl}}
\\
\bigskip

{\it Institute for Theoretical Physics,
University of Groningen\\
Nijenborgh 4, 9747 AG Groningen, The Netherlands}
\bigskip


{\bf C.M.~Hull}
\footnote{E-mail address: {\tt C.M.Hull@qmw.ac.uk}}
\\
\bigskip

{\it Physics Department, Queen Mary and Westfield College,\\
Mile End Road, London E1 4NS, U.K.}

\bigskip


{\bf Y.~Lozano}
\footnote{E-mail address: {\tt Y.Lozano@phys.uu.nl}}
\\
\bigskip

{\it Spinoza Institute, University of Utrecht,\\
Leuvenlaan 4, 3508 TD Utrecht, The Netherlands}

\vspace{1cm}


{\bf Abstract}
\end{center}
\begin{quotation}
\small

We discuss branes whose worldvolume dimension equals the
target spacetime dimension, i.e.~``spacetime-filling branes''. In addition to
the D9-branes, there
are 9-branes in the NS-NS sectors of both the IIA and IIB strings.
The worldvolume actions of these branes are constructed, via duality,
from the known actions of branes with codimension larger than zero.
Each of these types of branes is used in the construction of a string theory
with sixteen supercharges
by modding  out a type II string by an appropriate discrete symmetry and adding
32 9-branes.
These constructions are related by a web of dualities and each arises as a
different limit of the Ho\v rava-Witten
construction.

\end{quotation}

\vspace{1cm}

\newpage

\pagestyle{plain}


\newpage

\section{Introduction and Overview}

This paper is concerned with spacetime-filling branes and their role in the
construction of
string theories with 16 supersymmetries.
The most familiar case is that of the D9-brane of the IIB theory.
It is usually not consistent to consider a background with an arbitrary number
of D9-branes, but if
32  D9-branes are included and an orientifold is taken of the  IIB theory, then
the type I string
theory results.
In
\cite{Hull}, a number of other 9-branes were shown to arise.
Applying an   S-duality to the D9-brane gives a new 9-brane of the IIB theory
which carries a NS-NS charge and which
we will refer to as the NS-9B brane, and a T-duality takes this to a similar
NS-NS 9-brane of the IIA
theory, the
NS-9A brane. At finite string coupling, the IIA string should become  M-theory
compactified on a
circle, and the NS-9A brane and the D8-brane have a common M-theory origin,
which was postulated to be an
M9-brane (which is of course not a spacetime-filling brane). However, the
D8-brane arises in the
version of the IIA theory with a mass parameter, and the lifting of this to 11
dimensions involves
a number of complications, as we will discuss below. The charges for these
9-branes necessarily
arise in the superalgebras
\cite{Hull}, as we will review in section 2, giving further evidence that they
occur in the theory.

The low-energy effective action for a D9-brane is a Born-Infeld-type action for
the 10-dimensional
super-Yang-Mills theory. A spacetime-filling brane cannot move, of course, and
this is reflected in
the absence of any scalars in the effective action that could represent
translation zero-modes.
The action gives the gauge-field part of the type I effective action. In
section 4, we will
construct the   effective actions for the other spacetime-filling branes, which
will give the gauge
sectors of   the various  string theories with 16 supercharges; for example,
the NS-9B brane action
gives a Born-Infeld-type action with unusual dilaton dependence for the
Yang-Mills sector of the
SO(32) heterotic string. These actions also encode some other useful
information, such as the
dependence of the brane tension on the  moduli of the theory, as discussed in
section 3.

In section 5, we discuss how each of the spacetime-filling branes, together
with some of the
domain-wall branes, can be used
in the construction of superstrings with 16 supersymmetries, in the same way
that D9-branes are used
in the construction of the type I string. In each case, a type II string theory
is modded out by a
$\Z_2$ discrete symmetry in the presence of 32 spacetime-filling branes   to
give a new theory with 16 supercharges.
The   spacetime-filling branes give rise to the gauge symmetry in each case, as
branes that end
on the  spacetime-filling branes carry gauge charges.
For example, the SO(32) heterotic string is constructed by modding out the IIB
string by the S-dual
of the world-sheet parity symmetry in the presence of 32 NS-9B branes
\cite{Hull,Hnew}.
This puts the new spacetime-filling branes on a similar footing to the
D9-branes, and gives a
unified picture of the various  string theories with 16 supersymmetries.
 These constructions are lifted to M-theory in section 6, and it  is shown that
they arise as particular limits of the Ho\v rava-Witten construction \cite{HW},
giving an
M-theoretic justification for some of the assumptions made in section 5 and
\cite{Hull}.
Our conclusions are given in section 7.

We now return to the discussion of the IIA mass parameter and the M9-brane.
The D8-branes \cite{PW,B} are domain walls that divide regions with different
values of the mass parameter
of the {\it massive}  IIA string theory whose low-energy limit is the {\it
massive} IIA
supergravity of Romans \cite{Rom}.
In the massless case, the IIA string at finite coupling is M-theory
compactified on a circle whose
radius depends on the string coupling \cite{Wittt}, and it is natural to seek a
similar
interpretation of the massive theory.
In \cite{BLO}, it was shown that if the Romans supergravity is lifted to 11
dimensions, it must be
to a theory with explicit dependence on the Killing vector in the compact
direction.
Then
 11-dimensional covariance does not emerge
at
strong coupling if the mass is non-zero, and it is not clear if this  limit
exists.
This non-covariant theory does indeed have M9-brane solutions that give the
D8-brane on   double dimensional
reduction \cite{BLO,M9} and these give the NS-9A brane on vertical reduction.

Although the massive IIA   supergravity cannot be obtained from conventional
11-dimensional
supergravity, it was recently shown that the massive IIA string theory can be
obtained from
M-theory \cite{Hulm} by compactifying M-theory on a $T^2$ bundle over a circle
and taking a limit in
which all three 1-cycles shrink to zero size. The quantized mass parameter
arises as the topological
twist of the bundle. It seems plausible that the supergravity of \cite{BLO}
could emerge  as a limit
of this construction, with the explicit dependence on the Killing vector
emerging from the  M-theory background, not from any intrinsic modification of
M-theory.
 We will not discuss this further here, but some of our results give support to
the picture given
by the effective theory of
 \cite{BLO}.

\section{Branes and Charges}

Due to   recent developments in string theory,
it is by now well understood that there is an intricate relationship
between the \lq central'  charge structure of the spacetime
supersymmetry algebra and the spectrum of
BPS states that are described by supersymmetric
brane solutions \cite{ob1}. The generic rule is
that   a $p$-form   charge in D dimensions
 contains the charges for a $p$-brane and a ($D-p$)-brane
\cite{Hull,to1}\footnote{The spatial components  $Z_{i_1...i_p}$  of  a
$p$-form
  charge $Z_{M_1...M_p}$ give the
 charge carried by a
$p$-brane, whereas the    $Z_{0i_1\dots i_{p-1}}$ components
 can be dualized to give a  spatial ($D-p$)-form charge:
  $${\tilde Z}_{i_p\dots i_D}=\frac{1}{p!}
  \epsilon^{0i_1\dots i_{p-1}}_{\hskip 1.1truecm i_p\dots i_D}
  Z_{0i_1\dots i_{p-1}}\, ,$$
which is the charge carried by a ($D-p$)-brane.
The exceptions are a 0-form central charge, a self-dual central charge
 and the translation generator.
The BPS solutions carrying these are a 0--brane, a D/2--brane
and a gravitational wave, respectively.
These cases are special because the 0-form central charge has no time
component, the self-dual central charge has space components that are
not independent from
the time components, while the time component of the translation generator is
identified as the Hamiltonian.}.
This gives rise to the well-known BPS spectrum of type II string theory and
M-theory, together with the extra 9-branes discussed in the introduction; we
now review this, following \cite{Hull}.

The ten--dimensional IIA supersymmetry algebra with
central charges is given by ($\alpha = 1,\cdots ,32;\ M = 0,\cdots ,9)$:

\begin{eqnarray}
\{Q_\alpha, Q_\beta\} &=& \left (\Gamma^MC\right)_{\alpha\beta}P_M
+\left (\Gamma_{11}C\right )_{\alpha\beta} Z +\left (\Gamma^M\Gamma_{11}C
\right )_{\alpha\beta}
Z_M \nonumber
\w2
&& + {\textstyle {1\over 2!}}\left (\Gamma^{MN}C\right)_{\alpha\beta}
Z_{MN}
+{\textstyle{1\over 4!}}\left (\Gamma^{MNPQ}\Gamma_{11}C\right )_{\alpha\beta}
Z_{MNPQ}
\w2
&&+{\textstyle{1\over 5!}}\left (\Gamma^{MNPQR}\Gamma_{11}C\right
)_{\alpha\beta}
Z_{MNPQR}\, . \nonumber
\end{eqnarray}
Note that the right-hand-side contains the maximum number of allowed
central charges:

\begin{equation}
{\textstyle{1\over 2}} \times  32\times 33 = 1 + 10 + 10 + 45 + 210 + 252\, .
\end{equation}
Scanning the known IIA branes we find the following
correspondences between charges and BPS states:

\begin{eqnarray}
P_M &\rightarrow& {\rm W{\textstyle-}A}\, ,\nonumber
\w2
Z &\rightarrow& {\rm D0}\, ,\nonumber
\w2
Z_M &\rightarrow& {\rm NS{\textstyle -}1A\ and\ NS{\textstyle -}9A}\, ,
\nonumber
\w2
\label{susyIIA}
Z_{MN} &\rightarrow& {\rm D2\ and\ D8}\, ,
\w2
Z_{MNPQ} &\rightarrow& {\rm D4\ and\ D6}\, , \nonumber
\w2
Z_{MNPQR} &\rightarrow& {\rm NS{\textstyle -}5A\ and\ KK{\textstyle -}A}\, .
\nonumber
\end{eqnarray}
We find a gravitational wave (W--A), a
fundamental string (NS--1A), D$p$-branes ($p$ = 0,2,4,6,8), a solitonic
 five-brane
(NS--5A), a Kaluza-Klein monopole (KK--A) and a nine-brane (NS--9A).
All cases are well understood except for the NS--9A brane, which would be a
  spacetime-filling brane
in IIA string theory \cite{Hull}.

Another example is the ten-dimensional IIB supersymmetry
algebra with central charges. In this case there are two Majorana-Weyl charges
$Q_\alpha^i\ (i=1,2)$ with the same chirality. The algebra is given by

\begin{eqnarray}
\{Q_\alpha^i, Q_\beta^j\} &=&
\delta^{ij}\left ({\cal P}\Gamma^MC\right )_{\alpha\beta} P_M +
\left ({\cal P}\Gamma^MC\right )_{\alpha\beta} Z^{ij}_M  \nonumber
\w2
&&+ {\textstyle {1\over 3!}}
\epsilon^{ij} \left ({\cal P}\Gamma^{MNP}C\right )_{\alpha\beta} Z_{MNP}
+
{\textstyle{1\over 5!}}\delta^{ij}\left ({\cal P}\Gamma^{MNPQR}
C\right )_{\alpha\beta} Z^{+}_{MNPQR} \nonumber
\w2
&&+
{\textstyle{1\over 5!}}\left ({\cal P}\Gamma^{MNPQR}
C\right )_{\alpha\beta} Z^{+,ij}_{MNPQR}\, .
\end{eqnarray}
Here ${\cal P}$ is a chiral projection operator and $Z_M^{ij}\, ,
Z_{MNPQR}^{+,ij}$ are doublets of SO(2) (symmetric traceless representations).
The upper index $+$ indicates that the
charge is a self-dual
5-form. As in the IIA case
we have the maximum number of $p$-form charges:

\begin{equation}
{\textstyle{1\over 2}} \times  32\times 33 = 10 + 20 + 120 + 126 + 252\, .
\end{equation}
Scanning the known IIB branes we find the following
correspondences\footnote{
The D7-brane is not characterised uniquely by its 7-form charge, but
depends also on its $SL(2,
{Z})$ monodromy, and
one encounters various types of
7-branes in the literature (see e.g.~\cite{or1}). This is related to the
fact that for objects of co-dimension 2 an $SL(2, {R})$-transformation
keeps the charge invariant but changes the monodromy and the couplings to
strings and 5-branes \cite{Hull}.}:

\begin{eqnarray}
P_M &\rightarrow& {\rm W{\textstyle -}B}\, ,\nonumber
\w2
Z^{ij}_M &\rightarrow& {\rm D1\ and\ NS{\textstyle -}1B }\, ,\nonumber
\w2
&& {\rm D9\ and\ NS{\textstyle -}9B}\, ,\nonumber
\w2
\label{susyIIB}
Z_{MNP} &\rightarrow& {\rm D3\ and\ D7}\, ,
\w2
Z^{+,ij}_{MNPQR} &\rightarrow& {\rm D5\ and\ NS{\textstyle -}5B}\, ,
\nonumber
\w2
Z^+_{MNPQR} &\rightarrow& {\rm KK{\textstyle -}B}\, .
\nonumber
\end{eqnarray}
In this case we find a gravitational wave (W--B), a fundamental string
(NS--1B), D$p$-branes ($p$ = 1,3,5,7,9), a solitonic five-brane (NS--5B), a
Kaluza-Klein monopole (KK--B) and a further nine-brane (NS--9B).
We see that the IIB central charges suggest the existence of
two spacetime-filling branes: the D9-brane and the NS--9B brane.
The first one has been discussed in the context of D-branes
(see e.g.~\cite{pol1}),
and the second occurs in the work of \cite{Hull}.
All cases are well understood except for the NS--9B brane.

The last example to consider is the eleven-dimensional supersymmetry
algebra ($\alpha = 1,\cdots ,32; M=0,\cdots 10$):

\begin{eqnarray}
\{Q_\alpha, Q_\beta\} &=& \left (\Gamma^MC\right)_{\alpha\beta}P_M
+ {\textstyle {1\over 2!}}\left (\Gamma^{MN}C\right)_{\alpha\beta}
Z_{MN}  \nonumber
\w2
&&+{\textstyle{1\over 5!}}\left (\Gamma^{MNPQR}C\right )_{\alpha\beta}
Z_{MNPQR}\, .
\end{eqnarray}
Again, the algebra contains the maximum number of allowed central
charges:

\begin{equation}
{\textstyle{1\over 2}} \times  32\times 33 = 11 + 55 + 462\, .
\end{equation}
These central charges are related to the following M-branes:

\begin{eqnarray}
P_M &\rightarrow& {\rm W{\textstyle -}M}\, ,\nonumber
\w2
\label{susyM}
Z_{MN} &\rightarrow& {\rm M2\ and\ M9}\, ,
\w2
Z_{MNPQR} &\rightarrow& {\rm M5\ and\ KK{\textstyle -}M}\, .
\nonumber
\end{eqnarray}
We find a gravitational wave (W--M), a membrane (M2), a five-brane (M5),
a Kaluza-Klein monopole (KK--M) and a nine-brane (M9). For a recent
discussion of the M9-brane, see \cite{M9}. Note that
in this case we do not find any spacetime--filling branes.


\begin{figure}[h]
\scalebox{.5}
{
  \includegraphics{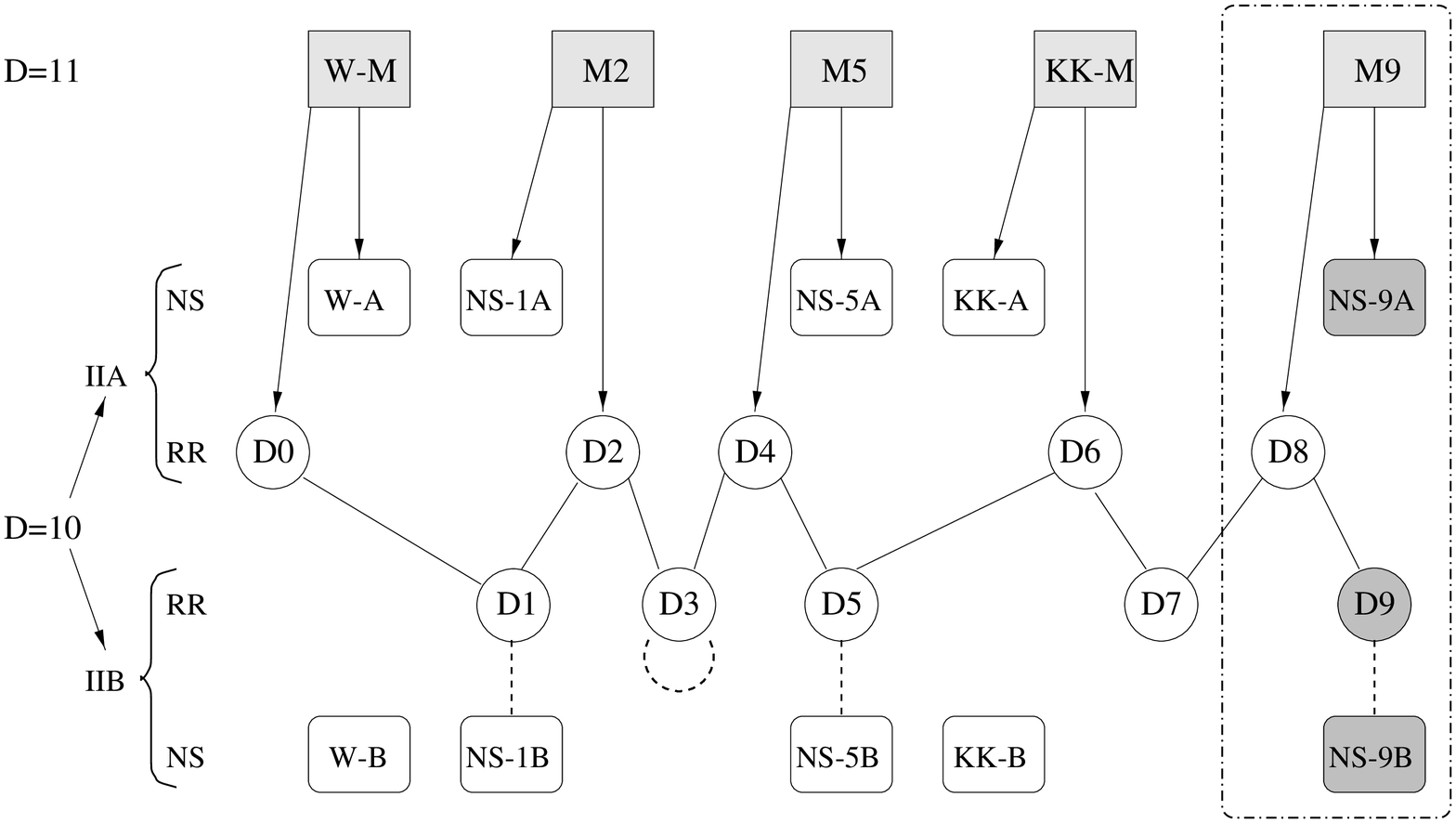}
}
\caption{{\bf Branes in ten and eleven dimensions.}
{\small
This figure depicts
the states of IIA, IIB and M-theories preserving half the supersymmetry. Each
carries a charge
occurring in the supersymmetry algebras (see eqs.~(\ref{susyIIA}),
(\ref{susyIIB})
and (\ref{susyM})). Arrows indicate dimensional reduction. In particular,
in this figure
vertical (diagonal) arrows indicate direct (double) dimensional
reduction. The solid (dashed) lines indicate T-duality (S-duality)
relations (the figure does not indicate all duality relations). The
five branes in the box at the right are the focus of this work.}}
\label{fig:scan}
\end{figure}


All branes mentioned above can be related to each other via T-duality,
S-duality and/or dimensional reduction (see Figure \ref{fig:scan}).
Formally, the D9-brane is obtained from other D-branes by T-duality, the
NS-9B brane is obtained from the D9-brane by S-duality, and   the NS-9A brane
of the IIA theory is obtained from the NS-9B brane by T-duality. Finally, the
M9-brane provides the 11-dimensional origin of both the D8-brane and the NS-9A
brane.
This is formal, as it is usually not consistent to have a single D9-brane, and
the same will apply to the other 9-branes related to this by duality. However,
it is  consistent to have 32 D9-branes together with an orientifold plane in
the construction of the type I string, and one of the purposes of this paper is
to give the analogous constructions involving the other 9-branes to give the
various
superstrings with sixteen supercharges, and show that they are related to the
type I construction by the appropriate S and T dualities. This we will do in
sections 5 and 6, but first we will discuss the effective actions and tensions
for these 9-branes.

\section{Effective Tensions}

In the next section we will construct the worldvolume actions
of the different spacetime--filling branes. It is instructive to first
consider the $p$-brane effective tensions, defined as the
mass per unit $p$-volume, which can be read off from the coefficient of the
leading Nambu-Goto term in the action.

In general the worldvolume actions of branes with $p$ spacelike dimensions
take the form:
\begin{equation}
S = {1\over l^{p+1} }\int d^{p+1}\xi\  e^{\alpha\phi} |k|^\beta \sqrt{ |{\rm
det} g|} +
\cdots\, ,
\label{wvaction}
\end{equation}
where the leading term is the Nambu-Goto term and the dots stand for
other worldvolume fields and the Wess-Zumino term.
Here $l$ is the fundamental length scale, which
is the (eleven-dimensional) Planck length $l = l_p$ in M-theory and
 is the string length
$l = l_s = \sqrt {\alpha^\prime}$ in string theory, in which case  we use a
(dimensionless) string-frame metric \footnote{
We use here  the same
conventions as \cite{ob1}. In the following sections we
will use the convention of, e.g., \cite{PW} where
one takes $l_s =1$ and works with
the  (string-frame) metric $g_{\mu\nu}$.}.
 For
M2- and M5-branes, $\alpha=\beta=0$, while for
conventional type II branes $\beta=0$ and $\alpha=-1$ for D-branes, $\alpha=0$
for the fundamental string and $\alpha=-2$ for NS-5 branes.

The KK monopole in $D$ dimensions it the product of  a self-dual Taub-NUT space
with $D-4$ dimensional Minkowski space and can be
considered as an extended object with $D-5$ spatial dimensions and one extra
isometry direction, the Taub-NUT fibre direction,  which is transverse to the
worldvolume.
In order to get the right counting of degrees of freedom this isometry is
gauged, such that the effective number of embedding scalars is 3, fitting
in a $D-4$ dimensional vector multiplet \cite{Hull}. The effective action of
the
monopole is   that of a $D-4$ dimensional gauged sigma model, with
Killing vector $k^\mu$ \cite{be1}.
For the M-theory KK monopole, the world-volume action is a gauged sigma-model,
with a target space isometry generated by a  (spacelike)
Killing vector $k^\mu$ being gauged \cite{be1}, and this leads to a term
$|k|^\beta$ in the world-volume action  \cite{be1}, where
$|k|^2 = -k^\mu k^\nu g_{\mu\nu}$ and $\beta=2$ (and $\alpha=0$).
This particular dependence with $|k|^2$ in front of
the action is needed in order to get, for D=11,
the correct $e^{-\phi}$ dilaton
coupling of the D6-brane after dimensional reduction.
The actions for the type II KK monopoles have
$\alpha=-2,\beta=2$ \cite{EJL}.

Given the worldvolume action (\ref{wvaction}) one can read off the
effective tension $\tau$:

\begin{equation}
\tau = {(g_s)^\alpha R^\beta\over l^{p+\beta +1}}\, ,
\end{equation}
where $R$ is the size of the Killing direction, $<|k|^2 >\equiv (R/l)^2$, and
$g_s = e^{<\phi>}$ is the string coupling
constant. Note that in string theory we have $\tau = \tau (l_s, g_s, R)$
while in M-theory we have $\tau =  \tau (l_p, R)$, since $\alpha = 0$.
Conventional branes have $\beta=0$, while the KK monopole has non-trivial
$R$-dependence, and more general objects with $R$-dependence have been found in
\cite{M9,H1,Hothers}.  For example,    the T-dual of a D6-brane is a circularly
symmetric D7-brane \cite{B}, i.e. one \lq smeared' over a circle, and the
S-dual of this is a 7-brane with $\tau \propto 1/g^3$.

In particular, the M9-brane has $\beta=3$ \cite{M9,H1}, and the
effective tensions of the other 9-branes then follow from this via dualities,
or can be
read  off  from the actions of the next section.
In Figure \ref{fig:tensions} we have given the effective tensions of
the three ten-dimensional
spacetime--filling branes. The same figure
also gives the effective tension of the ten-dimensional (eleven-dimensional)
domain wall D8 (M9).

It is instructive to verify
the S-- and T--duality
transformations of the effective tensions of the branes indicated
in Figure \ref{fig:tensions}.
The T-duality rules for $R, g_s$ and $l_s$ are:


\begin{figure}[h]
\scalebox{.63}
{
  \input{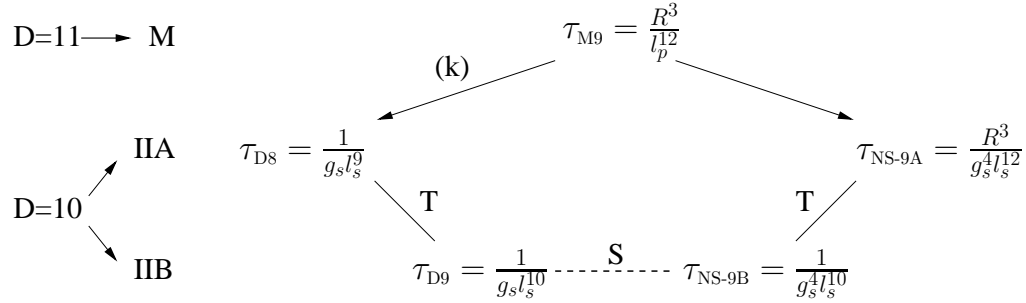}
}
\caption{{\bf Effective tensions.}
{\small
The figure gives the effective tensions of the three ten-dimensional
spacetime--filling branes (NS--9A, D9, NS--9B) and the ten-dimensional
(eleven-dimensional) domain wall D8 (M9). It also indicates
the different duality and reduction relations, using the same conventions
as   Figure \ref{fig:scan}.}}
\label{fig:tensions}
\end{figure}


\be
R^{\prime}= {l_s^2 \over R} \, ,\qquad g_s^{\prime}={g_s l_s \over R} \, ,
\qquad l_s^\prime = l_s\, ,
\label{T}
\ee
and the S-duality rules are given by

\be
R^{\prime}= R \, ,\qquad g_s^{\prime}={1 \over g_s} \, ,
\qquad l_s^\prime =  (g_s)^{1/2}l_s\, .
\label{S}
\ee
Finally, the reduction rules from D=11 to D=10 are given by

\begin{eqnarray}
l_p &=& g_s^{1/3}l_s\, ,\nonumber
\w2
R &=& \cases{l_s g_s \hskip .95truecm {\rm reduction\ in\ the\
R\ direction}\cr
             R \hskip 1.3truecm {\rm reduction\ in\ other\ direction}}
\label{reduction}
\end{eqnarray}

Using the rules (\ref{T}), (\ref{S}) and (\ref{reduction})
one can check that under T-, S-duality and reduction the
different effective tensions given in Figure \ref{fig:tensions}
correctly transform into each other.
It is understood that the T-duality relations are given by
$(R\tau_{\rm D9})^\prime = \tau_{\rm D8}$ and $(R\tau_{\rm NS-9B})^\prime
= \tau_{\rm NS-9A}$, respectively.

{}From Figure \ref{fig:tensions} we see that the NS--9A brane and the
M9-brane are
special in the sense that their effective tensions are proportional to $R^3$
and it is interesting to compare these with
KK monopoles which have $\tau \propto R^2$.
The KK monopole in $D$ dimensions has $D-5$ flat directions (which can be taken
to be $\R ^{D-5}$) and a Killing direction of radius $R$, and so can be thought
of as a $D-5$ brane with 3 transverse dimensions.
The NS--9A brane and the
M9-brane  each have 8 flat directions  (which can be taken to be $\R ^{8}$) and
a Killing direction of radius $R$ and so should be thought of as   8-branes.
In particular, we take $p=8$ in eq.~(\ref{wvaction}).
The NS--9A brane is space-filling while the M9-brane is a wall with one
transverse direction.

The KK-monopole cannot move in the
Taub-NUT isometry direction which, in order to give a charge to the monopole
is compact with finite radius $R$.  Taking the limit $R\to \infty $ of a
multi-monopole metric gives an ALE space.
For the NS-9A and M9  eight-branes,
the extra compact R direction can be decompactified
to give   nine-branes  only   for a certain number of coinciding
eight-branes (in combination
with an orientifold plane) but not for a single brane.

 The KK monopole
is a solution of the usual supergravity equations of motion,
but the M9-brane only occurs as a solution  of  supergravity   equations
modified by explicit dependence on a mass parameter and a Killing vector;  for
more details,
see \cite{M9,BLO}. One of the results of this work is that we
will clarify the special role of the Killing vector direction in the
case of the NS--9A brane and M9-brane.

The KK monopole with $D-5$ flat spatial dimensions and a circular  fibre
direction
carries a $D-5$ form charge \cite{Hull} while the NS-9A and M9-branes both have
8 flat spatial dimensions and a circular    direction and both carry a 9-form
charge. Thus in the KK monopole  case, the circular fibre is transverse to the
brane while in the NS-9A and M9-branes the circular direction is longitudinal,
so that the brane should be thought of as wrapping the circle.

\section{Worldvolume Actions}

\begin{figure}[h]
\scalebox{.5}
{
  \includegraphics{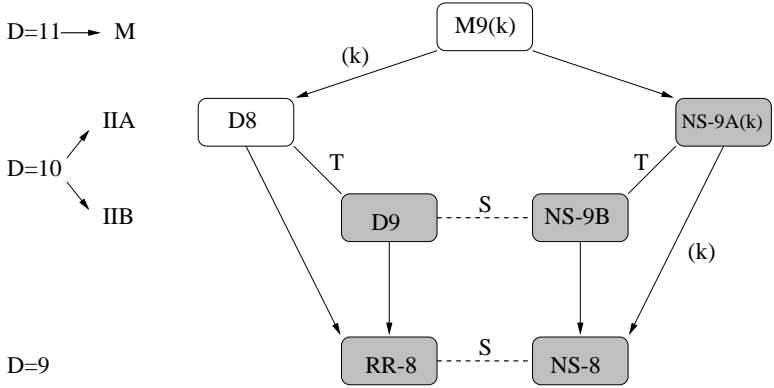}
}
\caption{{\bf Spacetime-filling branes in ten and nine dimensions.}
{\small The grey
areas indicate the (ten-- and nine--dimensional)
spacetime-filling branes whose actions are constructed in
this section.
To indicate the different reduction and duality relations,
we use the same notation as in Figure \ref{fig:scan}. Reduction over
a Killing isometry direction is indicated by the label (k).
The figure contains two more branes: the D8-brane or domain
wall in ten dimensions and
the M9-brane in eleven dimensions.
We have indicated that
the M9-brane and NS--9A-brane require the existence of a Killing vector $k$.
(See below).}}
\label{fig:branes}
\end{figure}

In this section we construct the worldvolume effective actions of
the different spacetime--filling branes.
Our starting point is the action for the D9-brane. By applying S-duality
we will obtain the action for the NS--9B brane, and from it
we will derive the action of the NS--9A brane through a T-duality
transformation.
The dimensional reduction
of the D9-brane and NS--9B brane leads to two spacetime-filling 8-branes
in nine dimensions:
the RR-8 brane and the NS-8 brane (see Figure \ref{fig:branes}).
All these branes will play a role
when we discuss in Section 5
the relation between spacetime--filling branes and
strings with sixteen supercharges.

Spacetime-filling branes are special in the sense that they have
zero codimension: there is no transversal direction. At first sight
this leads to two problems:

\begin{description}
\item{(1)} The lines of force cannot escape to infinity.

\item{(2)} The ``naive'' spacetime solution is flat Minkowski spacetime
since the harmonic function characterizing the brane solution only depends
on the transverse directions. Therefore the spacetime solution
does not appear to break any supersymmetry.
\end{description}

The D9-brane of the  Type IIB string theory \cite{Polchi}
is a brane carrying charge with respect to a
non-dynamical  RR 10-form potential, and $N$ of these lead to an open string
sector with $U(N)$ Chan-Paton factors.
This is inconsistent due to gauge anomalies and  charge non-conservation,
but if there are precisely 32 coincident D9-branes, then orientifolding by the
worldsheet parity reversal symmetry
$\Omega$ gives a consistent theory, which is the  Type I SO(32) string theory.
The orientifolding introduces
a space-filling  O9 orientifold fixed plane which has negative  charge
with respect to
the RR 10-form potential, and the role of the 32 D9-branes is to
cancel this  charge.

Similar issues arise for any $p$-brane in a space-time in which the transverse
space is compact, as the lines of force
again cannot escape to infinity. In general it is not consistent to put branes
in such a space-time, but it can be consistent
if the charge of the branes is cancelled by that of the fixed planes of some
discrete symmetry, or in some other way.
In general, then,  there are consistency conditions restricting which branes
can be placed in which space-times, and the case
of the D9-brane can be thought of as an example of this.

Nonetheless, it is sometimes convenient to think of a brane in an arbitrary
background.
In the actions for Dp-branes in arbitrary type II supergravity backgrounds,
the coupling to the NS-NS fields  is through a Born-Infeld kinetic term and to
the RR $n$-form gauge fields $C^{(n)}$ is through a
Wess-Zumino term.
In particular, this   gives an action for the D9-brane coupling to an arbitrary
  IIB supergravity background with fields \footnote{We include the SL(2)
  doublet of non-dynamical 10-form potentials: $B^{(10)}$ and $C^{(10)}$,
  to which the NS-9B brane and D9-brane couple, and the dual potentials.}:

\be
\label{IIBfields}
\left \{g \, ,\ \phi\, ,\ B^{(2)}\,  ,\
 B^{(10)}\, ,\ C^{(0)}\, ,\ C^{(2)}\, ,\
C^{(4)}\,  ,\ C^{(10)} \right \}\, .
\ee

Acting with S-duality   then gives the action for the NS-9B brane coupling to
the same set of fields, and
acting with T-duality using the rules of \cite{BHO} then gives the action for
the NS-9A brane coupling to the IIA
supergravity fields

\begin{equation}
\label{IIAfields}
\left \{g \, ,\ \phi\, ,\ B^{(2)}\,  ,\ B^{(10)} \,  ,\
C^{(1)}\, ,\
C^{(3)}\, ,\ C^{(9)} \right \}\, .
\end{equation}

In the orientifold construction, the IIB massless supergravity multiplet is
truncated to
the N=1 supergravity multiplet, with the following bosonic fields set to zero:

\be
\label{trunc:D9}
B^{(2)} = C^{(0)} = C^{(4)} = B^{(10)}= 0\, ,
\ee

\noindent and it is to this truncated background to which the
D9-brane of the type I theory couples. The Yang-Mills sector arises from
the world-volume gauge theory. Below, we will give the action for the type I
D9-brane
coupling to this truncated set of fields. We will give the abelian form of the
action, as
the formulation of a non-abelian
Born-Infeld action describing a  number of coincident D-branes is still an open
problem; one candidate for such an action is
given in \cite{tseytlin}.
Acting with S and T dualities then gives the N=1 truncations of the IIA and IIB
backgrounds to
which the other spacetime-filling branes will couple.

Since S-duality acts within the massless sector,
after applying S-duality, the NS--9B brane will couple to the S-dual of the
  truncation (\ref{trunc:D9}), given by

\be
\label{trunc:NS-9B}
C^{(2)} = C^{(0)} = C^{(4)} = C^{(10)}= 0\, .
\ee

The IIA supergravity background
(\ref{IIAfields})
  is truncated by:

\begin{equation}
\label{trunc:NS-9A}
C^{(1)} =  C^{(3)} = C^{(9)} = 0\, ,
\end{equation}

\noindent i.e. the T-dual of the projection (\ref{trunc:NS-9B}).
The NS--9A brane couples to this N=1 truncation.

Spacetime--filling branes in nine dimensions couple to N=1
truncations of the D=9 N=2 supergravity background. These truncations
follow via reduction from ten dimensions. Let us first introduce our
notation for the D=9 N=2 supergravity background.
We can consider
either its IIA or IIB origin, but it will be useful to consider both.

{}From IIB one gets the following D=9 N=2 background
fields\footnote{The notation in \cite{BHO} can be obtained after the
replacements: $A^{(1)}_\mu \rightarrow B_\mu$,
$B^{(2)}_{\mu\nu} \rightarrow B^{(1)}_{\mu\nu}$,
$B^{(1)}_{\mu} \rightarrow A^{(2)}_\mu$,
$C^{(2)}_{\mu\nu} \rightarrow B^{(2)}_{\mu\nu}$,
$C^{(1)}_\mu \rightarrow A^{(1)}_\mu$ and $C^{(0)} \rightarrow \ell$.}:

\begin{eqnarray}
g_{\mu\nu} &\rightarrow& \{g_{\mu\nu}\, ,\ A^{(1)}
\, ,\ \kappa\}\, ,\nonumber
\w2
\phi &\rightarrow& \phi ,\nonumber
\w2
B^{(2)} &\rightarrow&  \{B^{(2)}\, ,\ B^{(1)}\}\ ,\nonumber
\w2
B^{(10)} &\rightarrow& B^{(9)}\, ,\nonumber
\w2
C^{(0)} &\rightarrow& C^{(0)}\, ,
\w2
C^{(2)} &\rightarrow&  \{C^{(2)}\, ,\ C^{(1)}\}\, ,\nonumber
\w2
C^{(4)} &\rightarrow& C^{(3)}\, ,\nonumber
\w2
C^{(10)} &\rightarrow& C^{(9)} \, .\nonumber
\end{eqnarray}

\noindent
The RR-8 brane couples to the following N=1 truncation:

\be
\label{trunc:RR-8}
B^{(2)} = B^{(1)} = C^{(0)} = C^{(3)} = B^{(9)}= 0\, \,,
\ee

\noindent
since this is the reduction of the $\Omega$ projection (\ref{trunc:D9})
to 9 dimensions. The NS-8 brane couples to the reduction of
(\ref{trunc:NS-9B}):

\be
\label{trunc:NS-8}
C^{(2)} = C^{(1)} = C^{(0)} = C^{(3)} = C^{(9)}= 0\, \,.
\ee

We now indicate the IIA origin of the D=9 N=2 background fields since it
will be useful for the construction of the nine dimensional
spacetime-filling branes in subsection 4.2:

\begin{eqnarray}
g_{\mu\nu} &\rightarrow& \{g_{\mu\nu}\, ,\ B^{(1)}\, ,\
\kappa\}\, ,\nonumber
\w2
\phi &\rightarrow& \phi \, ,\nonumber
\w2
B^{(2)} &\rightarrow& \{B^{(2)}\, ,\ A^{(1)}\}\, ,\nonumber
\w2
B^{(10)} &\rightarrow& B^{(9)} \, ,
\w2
C^{(1)} &\rightarrow& \{C^{(1)}\, ,\ C^{(0)}\}\, ,\nonumber
\w2
C^{(3)} &\rightarrow& \{C^{(3)}\, ,\ C^{(2)}\}\, ,\nonumber
\w2
C^{(9)} &\rightarrow& \{C^{(9)}\, ,\ C^{(8)}\}\, .\nonumber
\end{eqnarray}

Since spacetime--filling branes have no transverse
directions their corresponding worldvolume actions do not
contain physical transverse embedding scalars. Indeed, all background
fields depend on the worldvolume embedding scalars only. For instance,
the dependence of the dilaton background field $\phi$ is given by

\be
\phi = \phi \left (X^\mu(\xi)\right )\, ,\hskip .5truecm
(\mu = 0,\cdots ,9)\, ,
\ee
where $X^\mu$ are the ten worldvolume
embedding scalars and $\xi$ are the ten
worldvolume coordinates. In the physical gauge

\be
\label{pg}
\vec X  = \vec \xi
\ee
we can identify the worldvolume of the spacetime--filling brane with the
target space and we obtain

\be
\phi = \phi(\vec\xi)\, .
\ee
We thus end up with a ten-dimensional field theory
consisting of a vector multiplet coupled to N=1
supergravity background fields.

In the next subsection we will first
construct the (bosonic part of)
the worldvolume actions of the three ten-dimensional
spacetime-filling branes: the D9-brane, the NS--9B brane and the NS--9A brane.
In the following subsection we will construct the worldvolume
actions of the two nine-dimensional spacetime--filling branes given in Figure
\ref{fig:branes}: the
RR-8 brane and the NS-8 brane, and show that they are related by an
S-duality transformation in nine dimensions.

\subsection{Actions for D=10 Spacetime--Filling Branes}
\subsubsection{The D9-brane action}

The world-volume theory for the spacetime--filling  D9-brane of the type II
theory
 is a 10-dimensional supersymmetric gauge theory with gauge field
$b$ coupling to the background fields (\ref{IIBfields}), while that of the type
I D9-brane
couples to the remaining fields after the truncation (\ref{trunc:D9}).
The (bosonic part of the)
worldvolume action of the type I D9-brane, in the physical gauge
(\ref{pg}),
is given by

\begin{equation}
S^{\rm (D9)} = S^{\rm (D9)}_{\rm DBI} + S^{\rm (D9)}_{\rm WZ}\, ,
\end{equation}
with Dirac-Born-Infeld (DBI) action:

\begin{eqnarray}
S^{({\rm D9})}_{{\rm DBI}} &=& -T_9 \int d^{10} \xi\ e^{-\phi}
\sqrt{|{\rm det} \left(g+ \alfa F \right)|} \nonumber
\w2
&=&
\label{D9}
- T_9 \int d^{10} \xi\ e^{-\phi} {\sqrt {|g|}}\biggl \{
1 - {\textstyle{1\over 4}}(\alfa)^2 {\rm tr}\ F^2 +
\w2
&& \hskip 2truecm
+{\textstyle{1\over 8}}(\alfa)^4\left [
{\textstyle{1\over 4}} ({\rm tr}\ F^2)^2 - {\rm tr}\ F^4\right ] +
\cdots \biggr\}\, ,\nonumber
\end{eqnarray}

\noindent where  the DBI field strength is

\be
\label{d9a}
F= 2\partial b \, ,
\ee

\noindent and the trace is taken over the spacetime
components,
so that e.g. ${\rm tr}\ F^2 = F_m{}^nF_n{}^m$.
The Wess-Zumino term is:
\begin{eqnarray}
S^{({\rm D9})}_{{\rm WZ}} &=& -T_9 \int 
\biggl\{
C^{(10)} +{\textstyle{1 \over 2!}} (\alfa)^2 {C}^{(6)}\wedge F\wedge F +
\w2
&& \hskip 2truecm
+ {\textstyle {1 \over 4!}}(\alfa)^4 {C}^{(2)}\wedge F\wedge F\wedge F\wedge F
+ \cdots \biggr\} \, .\nonumber
\end{eqnarray}

\noindent
Here $C^{(10)}$ is the RR 10-form,
${C}^{(6)}$ is the 6-form dual of $C^{(2)}$,
the dots indicate
gravitational contributions from the ``roof genus'' \cite{roof1,roof2}  and
we have used differential form notation\footnote{Our
  convention for a rank $r$ form is:
$$A_{(r)}=\frac{1}{r!}A_{(r)\mu_1\dots\mu_r}dX^{\mu_1}\wedge
\dots\wedge dX^{\mu_r}\, .$$}.

The cosmological term in the expansion of the DBI action
and the $C^{(10)}$ term in the Wess-Zumino part
are cancelled when one introduces the Orientifold O9 plane. The O9 plane
does not have any worldvolume dynamics but contributes, in the tadpole
cancellation, with
an energy density and charge opposite to that of 32 D9 branes, and so
corresponds to the following action:
\begin{equation}
S^{\rm (RR-O9)}= 32 \times T_9 \int d^{10} \xi \,\, \left \{
e^{-\phi} \sqrt{|g|}  +  C^{(10)} + \cdots \right \}\, .
\end{equation}
The dots represent other coupling terms, some of which have been
discussed in \cite{kn:orea}.

\subsubsection{The NS--9B action}

In order to obtain the worldvolume action of the NS--9B brane, we
apply an S-duality transformation to the D9-brane action.
The S-duality rules for the fields present in this action are:

\begin{equation}
\begin{array}{rcl}
\phi &\longrightarrow& -\phi \, ,
\\& &\\
g_{\mu \nu} &\longrightarrow& e^{-\phi} g_{\mu \nu} \, ,
\\& &\\
C^{(10)} &\longrightarrow& B^{(10)}\, ,
\end{array}
\label{Sdr}
\ee

\be
C^{(2)} \rightarrow B^{(2)} \, ,\qquad
{C}^{(6)} \rightarrow B^{(6)} \, ,
\ee

\noindent where $B^{(6)}$ is the 6-form dual to $B^{(2)}$.

\noindent The BI worldvolume 1-form $b$ transforms into a
1-form $c^{(1)}$ as follows:

\be
b \rightarrow -c^{(1)} \, ,\qquad c^{(1)} \rightarrow b \, .
\ee
Applying these rules to the action of the D9-brane we obtain
the following worldvolume action for the NS--9B brane:

\be
S^{\rm (NS-9B)} = S_{\rm DBI}^{\rm (NS-9B)} + S_{\rm WZ}^{\rm (NS-9B)}\, ,
\ee
with DBI term:

\begin{eqnarray}
S^{{\rm (NS-9B)}}_{{\rm DBI}} &=& -T_9\int d^{10}\xi\  e^{-4 \phi}
\sqrt{|{\rm det} \left( g - (\alfa) e^{\phi} F \right)|} \, ,\nonumber
\w2
&=&
\label{NS9B}
- T_9 \int d^{10} x\  {\sqrt {|g|}}\biggl \{
e^{-4\phi} - {\textstyle{1\over 4}}(\alfa)^2 e^{-2\phi}{\rm tr}\ F^2 +
\w2
&& \hskip 2truecm
+{\textstyle{1\over 8}}(\alfa)^4\left [
{\textstyle{1\over 4}}({\rm tr}\ F^2)^2 - {\rm tr}\ F^4\right ] +
\cdots \biggr\}\, ,\nonumber
\end{eqnarray}

\noindent
and  Wess-Zumino term:
\begin{eqnarray}
\label{NS9BWZ}
S^{({\rm NS-9B})}_{{\rm WZ}} &=& -T_9 \int 
\biggl\{
B^{(10)} +{\textstyle{1 \over 2!}} (\alfa)^2 B^{(6)} \wedge F\wedge F +
\w2
\hskip 2truecm
&&+  {\textstyle {1 \over 4!}}(\alfa)^4 B^{(2)} \wedge F\wedge
F\wedge F\wedge F
+ \cdots \biggr\} \, .\nonumber
\end{eqnarray}

\noindent
Here $F$ is the curvature of $c^{(1)}$:

\be
 F =2 \partial c^{(1)}\, .
\ee

As in the case of the D9-brane, we expect that
the cosmological term and the $B^{(10)}$ dependence of the WZ term
will disappear when we consider 32 coincident NS-9B branes and mod out by the
S-dual of $\Omega$, so that there is
a contribution from   the
S-dual of the RR O9 Orientifold (see section 5.3), which will carry $B^{(10)}$
charge and has action:
\begin{equation}
\label{9planeB}
S^{\rm (NS-O9)}= 32 \times
T_9 \int d^{10} \xi \,\, \left \{ e^{- 4\phi} \sqrt{|g|}  +  B^{(10)} + \cdots
\right \}\, .
\end{equation}


\subsubsection{The heterotic string effective action}

As we shall discuss in the next section, the heterotic
  SO(32) string is constructed by
modding out the IIB string by a discrete symmetry, introducing the
S-dual of the RR O9 Orientifold, and adding 32 NS-9B branes.
This then gives the Yang-Mills effective action of the heterotic
string to be
the sum of (the non-abelian form of)
the DBI action (\ref{NS9B}), the
   Wess-Zumino term (\ref{NS9BWZ})
and the
9-plane contribution  (\ref{9planeB}),
to give the action:

\begin{eqnarray}
S_{\rm het} &=& -T_9\int d^{10}\xi\  e^{-4 \phi}
\left[
\sqrt{|{\rm det} \left( g - (\alfa) e^{\phi} F \right)|}
- \sqrt{|{\rm det} g|}
\right]
\nonumber
\w2
&& -T_9 \int  \biggl\{
 {\textstyle{1 \over 2!}} (\alfa)^2 B^{(6)} \wedge F^2 +
  {\textstyle {1 \over 4!}}(\alfa)^4 B^{(2)} \wedge F^4
+ \cdots \biggr\}\nonumber
\w2
&=&
- T_9 \int d^{10} x\  {\sqrt {|g|}}\biggl \{
 - {\textstyle{1\over 4}}(\alfa)^2 e^{-2\phi}{\rm tr}\ F^2
\w2
&& 
+{\textstyle{1\over 8}}(\alfa)^4\left [
{\textstyle{1\over 4}}({\rm tr}\ F^2)^2 - {\rm tr}\ F^4\right ] +
\cdots \biggr\}\, ,\nonumber
\w2
&& -T_9 \int  \biggl\{
 {\textstyle{1 \over 2!}} (\alfa)^2 B^{(6)} \wedge F^2 +
  {\textstyle {1 \over 4!}}(\alfa)^4 B^{(2)} \wedge F^4
+ \cdots \biggr\}\, .\nonumber
\end{eqnarray}

It is interesting to compare with known results for the heterotic action.
A similar comparison between the D9-brane and the type I effective action
has been made in \cite{bachas}.
The WZ term involving $B^{(6)}$ gives the Chern-Simons coupling, the
WZ term involving $B^{(2)}$ gives the Green-Schwarz anomaly-cancelling term and
the first few terms of the expansion of the
DBI action agree with the results of \cite{tseytlin2}


\subsubsection{The NS--9A action}

We next apply a T-duality transformation on the NS--9B
brane (a T-duality on the D9-brane leads to the D8-brane).
As stressed in the caption of
Figure \ref{fig:branes}, we will see that this case is special in the sense
that the worldvolume action we obtain contains a Killing vector, which
effectively means that the action can only be defined in nine uncompactified
spacetime dimensions. Nevertheless, as will become clear later, it is
useful to consider this case on its own.

The NS--9A brane is in some ways like an  8-brane;
it has 8+1 non-compact directions, and there is one which is compactified.
It has an 8+1  dimensional world-volume, like an 8-brane, but the target space
has 10 scalars
$X^\mu$, 8+1 of which are identified with the world-volume coordinates in
physical gauge, and one of which is gauged.
T-duality  in this case can be understood through a double dimensional
reduction.
The target space is taken to have a circular direction $z$, say, with
$k=\partial / \partial z$ a
Killing vector, and so (in   physical gauge) the world-volume indices can be
divided into
$(i, \sigma)$ where $\sigma$ is the NS--9B brane worldvolume direction wrapping
the circle and $i$ runs over the remaining 9
dimensions.
 For the target space fields we apply the standard
T-duality rules  \cite{BHO}.
The T-duality
takes the NS--9B brane worldvolume vector field $c^{(1) }$ to a
9-dimensional vector $d^{(1)}_i$ and a scalar
$c^{(0)} $, which are the bosonic world-volume fields of the NS-9A brane.
The  T-duality  rules for the  worldvolume fields are given by

\be
\label{direc}
c^{(1) \prime}_i = - d^{(1)}_i \, ,\qquad c^{(1) \prime}_\sigma = - c^{(0)}
\, .
\ee

In order to explicitly calculate the T-dual of the NS--9B brane action
(\ref{NS9B})
it is convenient to use the basic matrix identity
(${\hat \imath}=(i,\sigma)$):
\be
{\rm det}A_{{\hat \imath}{\hat \jmath}} =
A_{\sigma \sigma} {\rm det} \left( A_{ij} - {1 \over A_{\sigma \sigma}}
A_{i \sigma}A_{\sigma j} \right) \, ,
\label{trick}
\ee
where the determinant on the right-hand-side is of one dimension
lower than the determinant on the left-hand-side.
Applying this identity to the worldvolume
action (\ref{NS9B}) we obtain for the determinant:

\begin{eqnarray}
|g_{zz}|\biggl |{\rm det}\biggl (
g_{ij} - {1 \over g_{zz}}g_{iz}g_{jz}
&-&(\alfa) e^\phi\bigl ( 2\partial_{[i} c^{(1)}_{j]} - 2\partial_{[i}
c^{(1)}_\sigma
g_{j] z}/ g_{zz}\bigr) \nonumber
\w2
&+&(\alfa)^2 {e^{2\phi} \over g_{zz}}\partial_i c^{(1)}_\sigma
\partial_j c^{(1)}_\sigma \biggr)\biggr| \, .
\end{eqnarray}
Using this intermediate result, we find that the (bosonic part of the)
NS--9A brane has
worldvolume fields (in the physical gauge (\ref{pg})):

\be
\{ d_i^{(1)}\, ,\ c^{(0)}\}\, ,
\ee
in terms of which the worldvolume action is given by

\be
S^{\rm (NS-9A)}   = S_{\rm DBI}^{\rm (NS-9A)}  + S_{\rm WZ}^{\rm (NS-9A)}\, ,
\ee with DBI term:

\begin{eqnarray}
\label{NS9A}
S^{\rm (NS-9A)}_{{\rm DBI}} &=& -T_8\int d^9 \xi \,\,
e^{-4 \phi} |k|^3 \times
\w2
&&\sqrt{|{\rm det} \left( \Pi - (\alfa)^2 e^{2\phi}
\partial c^{(0)} \partial c^{(0)} + \alfa |k|^{-1} e^\phi
{\cal H}^{(2)} \right)|} \, ,\nonumber
\end{eqnarray}

\noindent where $T_8 = \int d\sigma\ T_9$ and
$\Pi_{ij}$ is the reduced metric:

\be
\Pi_{ij}=\partial_i X^\mu \partial_j X^\nu ( g_{\mu \nu} +
|k|^{-2}k_\mu k_\nu) \, .
\ee

\noindent This action is invariant under translations generated by the
Killing vector $k^\mu$. We can define covariant derivatives:
$D_i X^{\mu}= \partial_i X^{\mu} + A_i k^{\mu}$, with $A_i$ a dependent
gauge field:
$A_i=|k|^{-2} \partial_i X^\mu k_\mu$, in such a way that
$\Pi_{ij}=D_i X^\mu D_j X^\nu g_{\mu\nu}$.
The field strength ${\cal H}^{(2)}$ is given by

\be
{\cal H}^{(2)} = H  - 2 (i_k B^{(2)}) \partial c^{(0)} \, ,
\ee

\noindent where $H=2 \partial d^{(1)}$.

The Wess-Zumino term reads

\begin{eqnarray}
&&S^{\rm (NS-9A)}_{{\rm WZ}} = -T_8  \int 
\,
\biggl\{ i_k B^{(10)} + {\textstyle {1 \over 2!}}(\alfa)^2
(i_k B^{(6)})\wedge H\wedge H +
\w2 \hskip -2truecm
&& -  (\alfa)^2 (i_k N^{(7)})\wedge H\wedge dc^{(0)}
- (\alfa)^2 (i_k B^{(6)})\wedge (i_k B^{(2)})\wedge
H\wedge dc^{(0)} + \nonumber
\w2 \hskip -2truecm
&&+ {\textstyle {1 \over 3!}} (\alfa)^4
\left( B^{(2)} -A \wedge (i_k B^{(2)}) \right)
\wedge H\wedge H\wedge H\wedge dc^{(0)} + \nonumber
\w2 \hskip -2truecm
&& + {\textstyle {1 \over 4!}} (\alfa)^4 A\wedge H\wedge H\wedge H
\wedge H \biggl\}
\, .\nonumber
\end{eqnarray}

\noindent The field $N^{(7)}$ is the Poincar{\' e} dual of the
Killing vector $k_\mu$ considered as a 1-form, and it was first encountered
in the action of the Kaluza-Klein monopole
\cite{BEL}. The T-duality transformation rules for $N^{(7)}$
and $B^{(6)}$ can be
found in \cite{EJL}. The field $B^{(10)}$ is the  NS 10-form potential
T-dual to the NS 10-form potential of the Type IIB theory.
Again, a cancellation of the cosmological
constant in the expansion of the DBI term and the
$i_k B^{(10)}$ term is supposed to take place when we consider 32
coincident NS-9A branes and we include the contribution of the
plane related by an S- and a T-duality to the   O9 orientifold plane of the
type
IIB theory.

The world-volume theory for the NS--9A brane is a 9-dimensional gauge theory
with a non-linear action for the
super-Yang-Mills multiplet with bosonic fields $\{ d_i^{(1)}\, ,\ c^{(0)}\}$,
coupled to 10 scalars $X^\mu$, nine of which
are eliminated in the physical gauge, and one of which drops out due to the
invariance under the isometry symmetry $X^\mu
\to X^\mu +\alpha k^\mu$, where $\alpha$ is an arbitrary local function on the
world-volume.
We see that the action of the NS--9A brane is a gauge theory coupled to a
gauged sigma-model and
contains a Killing vector.
The isometry is an essential ingredient, and the gauge symmetry means that the
collective coordinate for translations in the
Killing direction is absent, so that such translations are not physical.
 It
therefore requires an isometry direction and is effectively defined in
nine non-compact spacetime dimensions only.
Dimensional
reduction over the Killing   direction
gives the action of the NS-8 brane (see Subsection 4.2.2.).
We will see in Section 6 the reason
 why the 9-form central charge of the IIA supersymmetry
algebra cannot be realized by a proper ten-dimensional
spacetime--filling brane
whereas we have seen that this is possible for the two 9-form central
charges of the IIB supersymmetry algebra.

\subsection{Actions for D=9 Spacetime--Filling Branes}
\label{9}

In the previous subsection we have seen that the NS--9A brane
 has an 8+1 dimensional worldvolume
 and couples to background fields which break ten--dimensional
general covariance. The
dimensional reduction of the NS--9A brane over the Killing isometry direction
 leads to a spacetime--filling brane, which we
will refer to as the NS-8 brane,
 in nine dimensions.
Furthermore, by considering
the direct reduction of the D8-brane
we obtain a second  spacetime--filling brane, referred to as the RR-8 brane,
 in nine dimensions.
These two nine-dimensional spacetime--filling branes will play a role
when we discuss the relation between spacetime--filling branes
and strings with sixteen supercharges.
In this subsection we will construct their worldvolume actions
and show that they are related by an S-duality transformation.

\subsubsection{The RR-8 action}

To obtain the action of the RR-8 brane we perform a
direct reduction of the D8-brane. Note that the ten-dimensional
D8-brane is not a spacetime--filling brane but a domain wall. It therefore
couples to a IIA supergravity background. After direct reduction of the
D8-brane,
when the brane has become a spacetime--filling one,
 we perform the truncation (\ref{trunc:RR-8}).

Our starting point is the D8-brane action

\be
S^{\rm (D8)}  = S_{\rm DBI}^{\rm (D8)} + S_{\rm WZ}^{\rm (D8)}\, ,
\ee
with DBI term given by
\be
S^{({\rm D8})}_{{\rm DBI}} = -T_8 \int d^{9} \xi\ e^{-\phi}
\sqrt{|{\rm det} \left(g+ \alfa {\cal F} \right)|}\, ,
\ee
with
\be
{\cal F} = 2\partial b + {1\over \alfa} B^{(2)}\, ,
\ee

\noindent and Wess-Zumino term given by

\begin{eqnarray}
&&S^{\rm (D8)}_{{\rm WZ}}=T_8  \int 
\,
\biggl\{C^{(9)}-(2\pi\alpha^\prime) {C}^{(7)}\wedge {\cal F}
 \nonumber
\w2 \hskip -2truecm
&& +{\textstyle {1 \over 2!}}(2\pi\alpha^\prime)^2 {C}^{(5)}
\wedge {\cal F}\wedge {\cal F} -{\textstyle {1 \over 3!}}(\alfa)^3 C^{(3)}
\wedge {\cal F} \wedge {\cal F} \wedge {\cal F} \nonumber
\w2 \hskip -2truecm
&& +{\textstyle {1 \over 4!}}(\alfa)^4 C^{(1)}\wedge {\cal F}
\wedge {\cal F} \wedge {\cal F} \wedge {\cal F} \biggl\}
\, .
\end{eqnarray}

The direct reduction of this action
along the single transversal coordinate $Z$
gives, after performing the truncation
(\ref{trunc:RR-8}), the action of the RR-8 brane, which contains
the following worldvolume fields:

\be
\{ \ b_i\, ,\ Z \}\, .
\ee

\noindent Explicitly\footnote{
Now all target-space fields are nine-dimensional.},
\be
S^{\rm (RR-8)}_{{\rm DBI}} = -T_8 \int d^9 \xi\ e^{-\phi} \kappa^{-1/2}
\sqrt{|{\rm det} \left( g_{ij} - \kappa^2 \partial_i Z \partial_j Z+
\alfa {\cal F}_{ij} \right)|} \, ,
\label{RR8}
\ee
where the field strength ${\cal F}$ is defined as:
\be
{\cal F} = 2\partial b - {2 \over (\alfa)} A^{(1)} \partial Z \, .
\ee

\noindent The WZ term reads:

\begin{eqnarray}
&&S^{\rm (RR-8)}_{{\rm WZ}}=T_8  \int 
\,
\biggl\{C^{(9)}-(2\pi\alpha^\prime) C^{(6)} \wedge dZ
\wedge {\cal F}+
\w2 \hskip -2truecm
&&+{\textstyle {1 \over 2!}}(\alfa)^2 C^{(5)}\wedge {\cal F}
\wedge {\cal F} + \nonumber
\w2 \hskip -2truecm
&& -{\textstyle {1 \over 3!}}(\alfa)^3
\left(C^{(2)}-C^{(1)}\wedge A^{(1)} \right)
\wedge dZ\wedge{\cal F}\wedge {\cal F}\wedge {\cal F} + \nonumber
\w2 \hskip -2truecm
&& +{\textstyle {1 \over 4!}}(\alfa)^4 C^{(1)}
\wedge {\cal F} \wedge {\cal F} \wedge {\cal F} \wedge {\cal F} \biggl\}
\, . \nonumber
\end{eqnarray}

\noindent Here $C^{(9)}$ is the 9-form potential obtained after reducing
the D8-brane's 9-form $C^{(9)}$ and $C^{(6)}$ ($C^{(5)}$)
denotes the 6-form (5-form) dual to $C^{(1)}$ ($C^{(2)}$) in nine
dimensions.

\subsubsection{The NS-8 brane action}
Similarly, the dimensional
reduction of the NS--9A brane along the Killing isometry direction
gives the action of the NS-8 brane, with
worldvolume fields

\be
\{\ d_i^{(1)}\, ,\ c^{(0)}\}\, .
\ee

\noindent The action is given by

\be
S^{\rm (NS-8)} = S_{\rm DBI}^{\rm (NS-8)} + S_{\rm WZ}^{\rm (NS-8)}\, ,
\ee
with DBI term\footnote{The scalar $\kappa$ arises here from
$|k|^2= \kappa^2$.}:

\begin{eqnarray}
\label{NS8}
S^{\rm (NS-8)}_{{\rm DBI}} &=&
 -T_8 \int d^9 \xi e^{-4\phi}\kappa \times
\w2
&&
\sqrt{|{\rm det}\biggl(g_{ij} - (\alfa)^2 e^{2 \phi} \kappa
\partial_i c^{(0)} \partial_j c^{(0)}
+ (\alfa) \kappa^{-1/2} e^{\phi} {\cal H}_{ij} \biggr)|}
\, ,\nonumber
\end{eqnarray}
with ${\cal H}$ given by
\be
{\cal H}= 2 \partial d^{(1)} + 2 A^{(1)} \partial c^{(0)}
\equiv H + 2A^{(1)}\partial c^{(0)}\, .
\ee

\noindent The Wess-Zumino term reads:

\begin{eqnarray}
&&S^{\rm (NS-8)}_{{\rm WZ}}=-T_8 \int 
\,
\biggl\{B^{(9)}-(2\pi\alpha^\prime)^2 B^{(6)}\wedge H
\wedge dc^{(0)}+
\w2 \hskip -2truecm
&&+{\textstyle {1 \over 2!}}(\alfa)^2 B^{(5)}\wedge H
\wedge (H + 2A^{(1)} \wedge dc^{(0)})+ \nonumber
\w2 \hskip -2truecm
&& -{\textstyle {1 \over 3!}}(\alfa)^4 B^{(2)}\wedge H\wedge H
\wedge H\wedge dc^{(0)}-
{\textstyle {1 \over 4!}}(\alfa)^4 B^{(1)}\wedge H
\wedge H \wedge H \wedge H \biggl\}
\, . \nonumber
\end{eqnarray}

\noindent Here $B^{(9)}$ is the 9-form potential obtained after the
reduction of the 10-form potential $B^{(10)}$ of the NS--9A brane,
and $B^{(6)}$
($B^{(5)}$) is the 6-form (5-form) dual to $B^{(1)}$
($B^{(2)}$) in nine dimensions.

Note that both the RR-8 and the NS-8 branes contain one extra worldvolume
scalar: $Z$ and $c^{(0)}$, respectively. However, it is only in the case of the
RR-8 brane that this extra scalar can be used to oxidize the brane to a
conventional
brane in ten dimensions (the D8-brane). In the case of the NS-8 brane
it is not possible to do so in a D=10 covariant way. Instead one gets
a D=10 action with a Killing vector (the NS--9A brane).


Finally, we will show that the actions of the RR-8 brane and
NS-8 brane are S-dual (in nine dimensions) to each other.
The reduction of the S-duality rules (\ref{Sdr})
leads to the following S-duality rules for the nine-dimensional
target space fields \cite{BHO}:

\be
\begin{array}{rcl}
g_{\mu \nu} &\longrightarrow& \sqrt{\kappa} e^{-\phi} g_{\mu \nu} \, ,
\\& &\\
A^{(1)} &\longrightarrow& A^{(1)}\, ,
\\& &\\
\kappa &\longrightarrow& \kappa^{3/4} e^{\phi/2} \, ,
\\& &\\
e^{\phi} &\longrightarrow& \kappa^{7/8} e^{- 3\phi /4}  \, ,\\
\end{array}
\ee

\be
C^{(2)} \rightarrow -B^{(2)} \, ,\qquad
C^{(1)} \rightarrow B^{(1)} \, ,
\ee

\noindent and for their duals:

\be
C^{(5)} \rightarrow -B^{(5)} \, ,\qquad
C^{(6)} \rightarrow B^{(6)} \, .
\ee

\noindent The 9-form potentials transform as:

\be
C^{(9)} \rightarrow -B^{(9)} \, .
\ee

\noindent Furthermore, the
S-duality rules for the worldvolume fields are given by:

\be
Z \rightarrow -(2\pi\alpha^\prime) c^{(0)} \, ,\qquad
b \rightarrow d^{(1)} \, .
\ee
Applying these rules one may verify that the worldvolume action
of the RR-8 brane is S-dual (in nine dimensions) to the worldvolume
action of the NS-8 brane.
\bigskip
\bigskip

\section{String Theories, Orbifolds and Orientifolds}

In this section we will discuss how the spacetime-filling branes are used in
the construction of string theories
with 16 supercharges, and the way these are related by dualities.
In the previous section we   constructed the worldvolume actions of three
ten-dimensional (D9,
NS--9B and NS--9A) and two nine-dimensional (RR--8 and NS--8) spacetime-filling
branes and all
actions are related to each other via duality and/or reduction.
Orientifolding the Type IIB string theory by the worldsheet parity reversal
operator $\Omega$ of the
fundamental string (NS--1B brane) requires the addition of 32 D9-branes in
order to cancel the
anomalies and tadpole introduced by the O9-orientifold fixed plane and gives
the Type I SO(32) string theory.
However, the Type I SO(32) string theory is related,
via duality and/or reduction, to other string theories
with sixteen supercharges and this suggests that one might also be able to
describe
these other N=1 superstring theories by dividing out Type IIA or IIB string
theory by a
discrete symmetry, with the addition of
  a   set of spacetime-filling branes   in order to cancel the anomalies
introduced by the
projection.
In \cite{Hull,Hnew}, it was argued that the $SO(32)$ heterotic string can be
obtained by modding out
the IIB string with a (non-perturbative generalisation of) the operator
$(-1)^{F_L}$ (where $F_L$ is
the left-moving fermion number) in the presence of 32 NS-9B branes.
It is the purpose of this section to generalise this
to other cases and
identify
the perturbative symmetries of Type IIA/IIB string theories responsible for
the projections onto the various
string theories with sixteen supercharges.
In section 6,
we will
discuss  the (compactified) M-theory origin of
the Type II string theory discrete symmetries
and show how the constructions of this section all arise as limits of the
Ho\v rava-Witten construction.

A  unified description of the theories with 16 supercharges emerges.
The string theories with 16 supercharges
are   related by the dualities
 displayed
in Figure  \ref{fig:strings} and can be obtained from M-theory
compactifications.
We will refer to any duality between a theory and its strong coupling dual as
an S-duality, so that in this sense the S-dual of the Type I theory is the
SO(32) heterotic string.

\begin{figure}[h]
\scalebox{.5}
{
  \input{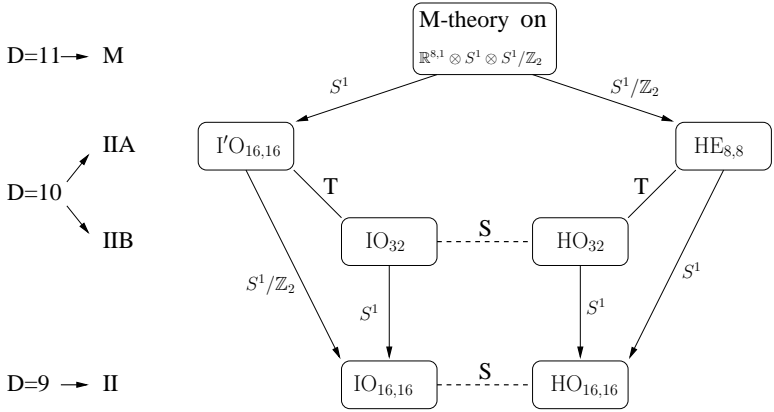}
}
\caption{{\bf String theories with 16 supercharges.}
{\small The duality and/or reduction relations between
theories that   result from $\Z_2$ projections of theories with 32 supercharges
(which are indicated on the left hand side of the figure). The M-theory origin
of two of the
theories with 16 supercharges is also indicated. The abbreviations used in  the
Figure are explained
in the text.}}
\label{fig:strings}
\end{figure}

This {\it defines} the discrete symmetries

\begin{equation}
\tilde \Omega = S \Omega S^{-1}, \qquad \hat \Omega = T \tilde \Omega T^{-1}
\end{equation}
while $I_{10} \Omega= T \Omega T^{-1} $ where $I_{10}$ is reflection in the
$x^{10}$
direction.
Each of these arise from a theory with 32 supercharges on modding out by
a $\Z_2$ symmetry, in a background with 16  9-branes or 8-branes (plus their 16
mirror images). The
type I   theory arises from modding out the IIB string by
the  world-sheet parity $\Omega$ with 32 D9-branes, and then the others arise
from
acting on these constructions with T and S dualities. The set of discrete
symmetries obtained in this way and used in the construction are shown in
Figure \ref{fig:z2}.

\begin{figure}[h]
\scalebox{.5}
{
  \input{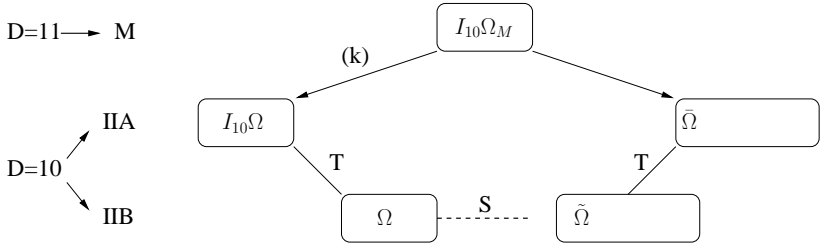}
}
\caption{{\bf $\mathbb{Z}_{2}$-symmetries.}
{\small
The $\mathbb{Z}_{2}$-symmetries of type II and M theories and their relations
under duality.
These are used in the constructions of the corresponding theory with 16
supersymmetries.
These symmetries are discussed further in
Subsections 5.3--5.5 and 6.1.}}
\label{fig:z2}
\end{figure}

\subsection{Orientifolds and Orbifolds}

An important ingredient in the discussion below is provided by the  orientifold
and orbifold
constructions. For an extensive review on both types of construction and
related issues, see
\cite{Dab}\footnote{For early references on orientifolds, see
\cite{orientifolds}.}.

Modding out a theory with a group of the form $G = G_{1} \cup G_{2} \Omega $
($G_1, G_2$ are
target-space symmetries and $\Omega$ involves an orientation--reversal on the
worldsheet) gives
either an orbifold (if $G_{2}$ is empty) or an orientifold (if $G_ {2}$ is
non-empty). For
consistency one also has to add a sector with twisted boundary conditions and
perform a projection
on  the twisted spectrum as well. In the orientifold-construction this usually
gives  an open-string
sector, i.e. strings which are
 closed modulo $\Omega$.
Here we will only consider groups $G=\bb{Z}_2$.

For orientifolds, the $p$-planes that are invariant under the orientation
reversal $\Omega$ are called
(perturbative) orientifold-fixed O$p$--planes. They couple to a RR--potential
and are negatively
charged. When the O$p$--planes have no non-compact transverse direction their
charge must be
cancelled. This is achieved through the addition of a fixed number of
D$p$--branes, with opposite
charge with respect to the same RR--potential, on which the open strings can
end. They provide the
Chan-Paton factors for the open strings.
For other (non-orientifold) groups $G$, anomalies are cancelled and charges
conserved
if other types of branes are added, such as the NS-9 branes, as
we will see. When such \lq background' D-branes or other branes are added,
$p$-branes that end on
these background branes carry generalised Chan-Paton indices and the action of
$G$ is extended to
act on these indices.

If $G$ is a symmetry of a  perturbative string theory $X$, this construction
leads to a new
perturbative string theory $X/G$, provided the correct set of
anomaly-cancelling branes is added.
Acting with T-duality then relates this to other perturbative constructions.
The question   arises as to whether the  construction of $X/G$ from $X$ can be
extended to the
non-perturbative  string theory.
First of all, this requires that $G$ should be a symmetry of the full
non-perturbative generalisation of $X$.
If it is, then we can consider modding out the full theory by $G$ (adding a
twisted sector and
projecting onto invariant states) and this should give a non-perturbative
generalisation of
$X/G$.
In particular, we can then take the strong coupling limit in which  $X$ is
replaced by a dual
theory $\tilde X$ which is a perturbation theory in $\ti g=1/g$, where $g$ is
the coupling of $X$.
(For example, if $X$ is the IIA string, $\ti X$ is M-theory on a circle of
radius $R$, with coupling
constant $\ti g=l_s/R$.)
Then the original symmetry extrapolates to a symmetry $\ti G$ of the dual
theory $\ti X$, and
modding out $\ti X$ by $\ti G$ to give $\ti X/\ti G$, which should be
the strong coupling limit of $X/G$,
this will be a perturbative construction (in $\ti g$) if $\ti G$ is a
perturbative symmetry of
$\ti X$; this will be the case for the examples considered here. If the
original construction
of
$X/G$ required the addition of  background branes, then the construction of
$\ti X/\ti G$ will require the addition of the branes that are the strong
coupling extrapolation of
these.
If $X/G$ is an orientifold construction, then its strong coupling dual
$\ti X/\ti G$ often turns out to be an orbifold construction, so that
non-perturbatively it doesn't
  make sense to differentiate between orientifolds and orbifolds, and   we
shall simply refer to
modding out by a discrete symmetry.
Note that in some cases, different results arise depending on whether or not
background branes are
added.

For example, if $X$ is the IIB string and $G=\Omega$, then $X/G$ with 32
D9-branes gives $X/G$ as
the type I string. The strong coupling limit is given by a dual perturbative
IIB string theory in
which the symmetry acts through $\ti G=(-1)^{F_L}$. Then $\ti X/\ti G$ with 32
NS-9B branes gives
the $SO(32)$ heterotic string, which is indeed the strong coupling limit of the
type I string.
However, in this case one can also consider modding out the IIB theory by
$(-1)^{F_L}$ without
adding any 9-branes, and it has been argued that this gives the IIA string
\cite{Dab}, at least
perturbatively.
In what follows, we will discuss these and other examples in more detail.

\subsection{D9/Type I SO(32)}

This case is the one that is best understood (for a review, see
e.g.~\cite{pol1}) and we have
discussed it already in different parts of the paper.
The operator $\Omega$ acts on the perturbative IIB theory through the
worldsheet parity reversal
of the fundamental string:

\be
\Omega:\hskip .5truecm \sigma \longrightarrow \pi - \sigma\, .
\label{o}
\ee

In orientifolding by $\Omega$, it is necessary to add
32 coincident D9-branes   to cancel the
negative RR-charge of the O9 orientifold fixed plane.
The invariant sector is the unoriented Type I closed string theory
and there is a
\lq twisted sector' of open strings (i.e.~strings that are closed modulo an
$\Omega$ transformation)
whose Chan-Paton factors are provided by the 32   D9-branes. The SO(32) gauge
symmetry arises after
the projection of the U(32) worldvolume gauge group of the 32 D9-branes.
For the massless bosonic fields, restricting to the invariant sector gives the
 truncation (\ref{trunc:D9}) of the
IIB supergravity theory, while the open strings give the SO(32) gauge fields.

\bigskip

\subsection{D8/Type ${\rm I}^\prime$  }

This case has also been discussed in the literature (see e.g.~\cite{pol1}).
It can be obtained, via T-duality, from the previous case compactified on a
circle.
The T-duality takes the IIB theory to the IIA string theory and the operator
$\Omega$ to

\be
T\, \Omega\, T^{-1} = I_{10}\ \Omega\, ,
\ee
 where $I_{10}$ is the
spacetime reversal of the compact $x^{10}$ direction
and   $\Omega$   is the worldsheet parity reversal operator acting on the
fundamental string of
the IIA or IIB theory.
Note that $\Omega$ is not a symmetry of the IIA string theory, but $ I_{10}\
\Omega$ is, acting as

\begin{equation}
I_{10}\Omega:\hskip .5truecm  x^{10} \longrightarrow - x^{10}\, ,
\hskip .5truecm \sigma \longrightarrow  \pi - \sigma\, .
\label{i9o}
\end{equation}

The type ${\rm I}^\prime$ theory is defined by orientifolding the IIA string by
 $ I_{10}\ \Omega$.
  As the compact
direction  is divided  out by the $I_{10}$ operator, the Type
${\rm I}^\prime$ theory is defined over the manifold $\R^{8,1} \times
S^1/\mathbb{Z}_{2}$. There are
  two O8 planes (each with RR charge --16) sitting at the two
fixed points  of the $\mathbb{Z}_{2}$, $x^{10}=0,\pi$.
To cancel the RR charge, it is necessary to add 16 D8-branes (and their 16
mirror images), and these can be located   anywhere
on  the circle.
The space $ S^1/\mathbb{Z}_{2}$ can be viewed as the interval $0\le x^{10}\le
\pi$, so the
construction gives the IIA string compactified on the interval with  16
D8-branes that can be
located anywhere  on  the interval.
This orientifold gives  the truncation (\ref{trunc:RR-8}) of
the massless bosonic sector of the Type IIA string theory.

The type I string compactified on a circle gives a 9-dimensional theory with
gauge group
$SO(32)\times U(1)\times U(1)$ (or $SO(32)\times SU(2)\times U(1)$ at the
self-dual radius)
and this is T-dual to the type ${\rm I}^\prime$ theory with the same gauge
group.
However, Wilson lines can be introduced in the compactification of the type I
string and by
choosing these the gauge group can be made to be $G_{17}\times U(1)$ where
$G_{17}$ is any
simply-laced rank-17 gauge group, and for any choice there is  a type ${\rm
I}^\prime$ dual with the same gauge
group.
 The 16 parameters determining the Wilson lines in the type I picture
correspond to the 16
positions of the D8-branes in the type ${\rm I}^\prime$ theory.
The type I momentum $p^9$ and D-string winding number in the $x^{10}$ direction
correspond in the type
${\rm I}^\prime$ picture to the fundamental string winding number in the
$x^{10}$
direction and the D0-brane charge.

The relation between the positions of the D8-branes and the gauge symmetry has
been discussed in
\cite{BGS} (an alternative description of the symmetry enhancement is given in
\cite{Sei}). If there
are 8  coincident
D8-branes (and their mirrors) at each O8-plane then the gauge group is
$SO(16)  \times  SO(16)\times U(1) \times U(1)$. If one of the D8 branes is
displaced from each of
the O8-planes by a distance $d$, the generic configuration has $SO(14)  \times
SO(14)\times U(1)^4$
symmetry, but if the displacement $d$ takes a certain critical value, this is
enhanced to
$E_8  \times  E_8\times U(1) \times U(1)$, and then if the size of the
$S^1/\mathbb{Z}_{2}$ is tuned to be twice this displacement, so that the two
displaced branes are
coincident, the group is enhanced to $E_8  \times  E_8\times SU(2) \times
U(1)$.
If all the D8-branes are at the same O8-plane, then the symmetry is generically
$SO(32)\times U(1)^2$, and if the coupling constant is tuned to a critical
value, this is enhanced
to $SO(34)\times U(1) $.
Bound states of D0-branes with O8-planes  play a crucial role in these symmetry
enhancements \cite{Lowe,Sil,BGS}.

The D8-branes are domain walls that divide regions with different
values of the mass parameter
of the {\it massive}  IIA string theory whose low-energy limit is the {\it
massive} IIA
supergravity of Romans \cite{Rom}.\footnote{The explicit D8-brane solution has
been given in
\cite{PW,B}.} In the SO(16) $\times$ SO(16) configuration the 8 D8-branes at
each orientifold plane
cancel  the RR charge of the O8-plane locally and the bulk theory is the usual
massless IIA string
theory, while in all other cases, the charges are only cancelled globally and
the massive IIA string
is the bulk theory in at least some of the regions between branes.
 At strong
coupling the Type ${\rm I}^\prime$ string theory can be described in terms of
(compactified)
M-theory where an eleventh direction (with coordinate $x^{11}$) has been
developed.
The coupling constant $g_{I^\prime}$ of the Type ${\rm I}^\prime$ theory
is given by

\begin{equation}
\label{qwerty}
g_{I'}=R_{11}^{3/2} \, ,
\end{equation}

\noindent where $R_{11}$ is the compactification radius of the $x^{11}$
coordinate.
Only for the SO(16) $\times$ SO(16)
configuration can the limit $R_{11} \rightarrow \infty$  be defined \cite{PW};
this will be discussed
further in Section 6.

Note that although a D8--brane in $\R^{9,1}$ is a domain wall,
when it is embedded in a manifold $\R^{8,1} \times S^1/\Z_2$ it has no
noncompact transverse
direction, and therefore effectively behaves like a spacetime--filling brane.
Reduction along the $
S^1/\Z_2$-direction gives an RR--8 brane, filling the whole $\R^{8,1}$
spacetime. These
spacetime--filling branes couple to the truncation (\ref{trunc:RR-8}) of the
massless D=9 Type II
background fields,  and provide Chan-Paton factors to the open strings of the
D=9 type I string
theory.
For example, for the SO(16) $\times$ SO(16)  configuration, the branes provide
 SO(16) $\times$
SO(16)  Chan-Paton factors and give rise to the 9-dimensional theory with gauge
group
SO(16) $\times$ SO(16) $\times$U(1)$\times$U(1), which we will refer to as the
$IO_{16,16}$ theory.
 The connection between this theory and the other N=1 theories is
indicated in Figure \ref{fig:strings}.

\subsection{NS--9B/Heterotic SO(32)}

This case has been discussed in \cite{Hull}, and more extensively in
\cite{Hnew}, and arises from
taking the strong coupling limit of the orientifold construction of the type I
theory.
Acting with the SL(2,\Z) transformation $S$ that takes weak to strong coupling
takes the IIB theory to itself
\footnote{The S-duality transformation takes the full non-perturbative
IIB theory to itself, but takes the perturbative IIB string theory defined by
perturbation theory in
$g$ to the   dual string theory defined by perturbation theory in $\ti g =1/g$;
the two dual string
theories are equivalent, but arise as different \lq slices' of the full
non-perturbative theory.},
and   takes
$\Omega$ to a symmetry

\be
S \Omega S^{-1} =  \tilde \Omega
\ee
This  acts as the   perturbative symmetry

\be
  \tilde \Omega = (-1)^{F_L}
\ee
on perturbative IIB string states (where $F_L$ is the  left--moving  fermion
number operator), and
reverses the parity of the D-string world-sheet.
The S-duality takes the 32 D9-branes to 32 NS-9B branes and as the S-dual (i.e.
strong-coupling
limit)  of the type I string is the SO(32) heterotic string, modding out
the IIB string by $\ti \Omega$ in the presence of
32   NS-9B branes should give the SO(32) heterotic string \cite{Hull,Hnew}.
Projecting onto the $\ti \Omega$-invariant sector projects the IIB supergravity
multiplet onto the
N=1 supergravity multiplet with the
 truncation (\ref{trunc:NS-9B}) of the massless bosonic fields.
In \cite{Hnew}
it is shown that the gauge structure of the heterotic string emerges as
follows.
There are open \lq D-strings' (arising from the IIB  D-strings) which end on
the 32
NS-9B branes and so carry SO(32) Chan-Paton factors. There are also open
D-strings
which have one end on the fundamental string and the other on the
NS-9B brane, so that they tether an SO(32) charge to the fundamental string.
Some of the segments of the \lq fundamental' string will in fact be (1,p)
strings for various values of $p$, as follows from charge conservation.
For example, attaching a D-string and an anti D-string to a fundamental string
leads to a    \lq fundamental' string
split into two segments by the junctions and which is a (1,0) string for one
segment and a (1,1) string for the other segment.
In the weak coupling
limit, the tension of these D-strings becomes infinite, so that the SO(32)
charges are pulled onto
the heterotic string world-sheet  give rise to the SO(32)
current algebra and the gauge sector of
the theory, and at the same time the (1,p) segments
of the string also contract, leaving a fundamental string; see [2] for further
details.


Thus, the Heterotic SO(32) string theory
arises by modding out the
 Type IIB string theory by $\tilde \Omega$.

This can be generalised to the case of (p,q) 9-branes. The (p,q) 9-brane is
obtained from the
D9-brane via an $SL(2,\Z)$ transformation $S_{p,q}$ and this takes $\Omega$ to

\be
S_{p,q} \Omega S_{p,q}^{-1} =  \tilde \Omega_{p,q}
\ee
and leads to a construction in which the IIB theory is modded out by $ \tilde
\Omega_{p,q}
$ with 32 (p,q) 9-branes.

\subsection{NS--9A/Heterotic }

This case has not been discussed in the literature,
although  it is
related to known cases.  Our starting point is the construction in section 5.4
of the SO(32)
heterotic string by modding out the IIB string with $\tilde \Omega$ in the
presence of 32 NS-9B
branes. Compactifying on a circle and performing a T-duality  (without Wilson
lines),  the IIB string
becomes the IIA string,  the SO(32) heterotic string is mapped to itself, the
NS-9B branes are
mapped to NS-9A branes  and $\tilde \Omega$ is mapped to

\be
T \tilde \Omega T^{-1} = \hat \Omega
\ee
where $\hat \Omega
$ is the symmetry which acts on perturbative IIA string states
through the perturbative symmetry

\be
T (-1)^{F_L} T^{-1} = (-1)^{F_L^c}\, ,
\label{TLC}
\ee
where $F_L^c$ is the left-handed fermion number of the IIA string.
This then gives a construction of the SO(32) heterotic string compactified on a
circle
from the type IIA string compactified on a circle modded out by $ \hat \Omega
$ with 32 NS-9A branes. In particular, it yields the  truncation
(\ref{trunc:NS-9A})
of the massless bosonic sector of the IIA string.
The open D-strings of the IIB construction that give rise to the gauge
sector of the heterotic strings   become, after the T-duality, D0-branes and
open
D2-branes.

For the heterotic string compactified on a circle, one can obtain the  gauge
group
$G_{17} \times U(1)$ for any simply-laced rank-17 group $G_{17}$ by adding
Wilson lines, and the question arises as to what these Wilson lines correspond
to in the dual picture;
they cannot correspond to   positions of the NS-9A branes as they are
space-filling (and in the
perturbative  IIA string, there is no extra dimension in which to displace
them). In the IIB
picture,  the NS-9B brane has world-volume vector fields $c^{(1)}$ taking
values in the SO(32) Lie
algebra, and the Wilson lines correspond to giving expectation values to the
component
$c^{(1)}_\sigma$ in the compact direction which take values in a Cartan
sub-algebra of SO(32). The
NS-9A brane world-volume theory is, after fixing the Killing symmetry $\delta
X^\mu=\alpha k^\mu$, a
9-dimensional super-Yang-Mills multiplet (with a vector field and Lie
algebra-valued scalar
$c^{(0)}$) coupled to a supergravity background, with a non-linear DBI action.
After the T-duality transformation (\ref{direc}), the Wilson lines correspond
to giving expectation values to
the scalars
$c^{(0)}$ that arise in the NS-9A brane world-volume theory, where the
expectation values are   in
a Cartan subalgebra.

Dimensionally reducing to 9 dimensions, this gives a construction of the
9-dimensional heterotic
string theory from the 9-dimensional type II string with 32 of the
spacetime--filling NS-8 branes
discussed in Subsection 4.2.2.
These branes
couple  to the truncation (\ref{trunc:NS-8}) of the D=9  Type II  massless
background fields.

In the next section, we will lift  this construction to M-theory, and show that
the expectation values of the scalars
$c^{(0)}$ correspond to displacing the M9-branes in an 11'th dimension which is
an $S^1/\Z_2$.

\subsection{Orbifolding with $(-1)^{F_L}$}

Orbifolding the perturbative IIB string with $(-1)^{F_L}$, in the absence of
any 9-branes, gives the
IIA string \cite{Dab}.
In the untwisted sector, all R-R and R-NS states of the IIB string are
projected out, and
the twisted
sector introduces states that make up the R-R and R-NS sectors of the IIA
string.
In the massless
sector, the IIB supergravity can be decomposed into an N=1 supergravity
multiplet and a right-handed
N=1 gravitino multiplet, and in the untwisted sector this gravitino multiplet
is projected out.
The twisted sector introduces a left-handed
N=1 gravitino multiplet, which combines with the N=1 supergravity multiplet to
give the IIA
supergravity multiplet.
The orbifolding of the IIB string to obtain the IIA string is a conformal field
theory
construction, and it is interesting to ask whether it can be extended to a
non-perturbative
construction that works for finite coupling or strong coupling.

In subsection 5.4, we  considered modding out the IIB string by the same
symmetry, $(-1)^{F_L}$, but
in a background with 32 NS-9B branes, and   this gave the SO(32) heterotic
string.
Thus it appears that modding out by the same symmetry can give different
results, depending on
whether or not 9-branes are added.
The perturbative symmetry $(-1)^{F_L}$ extends to the symmetry $\ti \Omega$ of
the full IIB
theory and at strong coupling becomes the world-sheet parity operator $\Omega$
of the dual
weakly-coupled IIB string theory, while the NS-9B branes become the D9-branes.
Modding out by $\ti \Omega$ with 32 NS-9B branes to give the heterotic string
is then S-dual to the
modding out by $\Omega$ with 32 D9-branes to give the type I string.
Suppose that the construction without 9-branes of the IIA from the IIB  extends
to the full
non-perturbative theory. Then applying an S-duality would give  the modding out
of the IIB string by
$\Omega$, without any D9-branes, to give the strong-coupling limit of the IIA
string, which is
M-theory. The untwisted sector gives an $N=1$ 10-dimensional supergravity
multiplet, and  the absence
of D9-branes means that there are no open strings. The twisted sector must
provide the extra states
needed to give M-theory.

Thus modding out the IIB string  by $\ti \Omega$ gives different results,
depending on whether or
not NS-9B branes are added.
With 32 NS-9B branes, this gives the SO(32) heterotic string and, as we shall
see in more detail in
the next section, extends to the full non-perturbative theory.
Without any 9-branes, no gauge group is introduced and  this takes the
perturbative IIB string to the
 perturbative IIA string. Either this does not extend to the full
non-perturbative theory, in which
case  it would be interesting to understand what goes wrong, or it does extend,
in which case
it would have an S-dual construction of M-theory on a large circle of radius
$R$ from orientifolding
the  IIB string with $\Omega$ at small coupling $g_B = l_s/R$, and this
would be interesting to understand directly.

Similarly, the IIA string can be orbifolded by $(-1)^{F_L^c}$
to give the IIB string. If this can be extrapolated to strong coupling, the
S-dual construction
would be as follows. The IIA string becomes M-theory compactified on a large
circle of radius $R=l_s
g_A$ at
strong string coupling $g_A$, and the symmetry $(-1)^{F_L^c}$ extends to the
M-theory symmetry used in the Ho\v rava-Witten construction \cite{HW} (we
refer to this
symmetry as $\Omega _M I$ in the next section). Then modding out M-theory with
this symmetry would
give a Ho\v rava-Witten-type construction, but without any gauge sectors
introduced at the fixed
points of the symmetry, and would give the weakly-coupled IIB string with
coupling $g_B = l_s/R$.
Again, it would be interesting to investigate this further.

\bigskip
\section{Relation with M-Theory}
\subsection{The Ho\v rava-Witten picture}

In this section, we will show how the constructions of string-theories with 16
supersymmetries
 of the
last section can all be lifted to M-theory, and that they all arise as
particular limits of the
Ho\v rava-Witten picture of M-theory compactified on $\R^{8,1}\times S^1\times
S^1/\Z_2$ \cite{HW}.
This will give an M-theoretic justification of some of the assumptions made in
section 5,
 allow us to discuss the non-perturbative generalisations of the constructions,
and give insights
into the M-theoretic origin of D8-branes.

Consider first M-theory compactified on a 2-torus with radii $R_{10},R_{11}$.
When one of the radii is large and the other small, the theory is described by
a weakly coupled IIA
string theory; for example, if $R_{11}$ is small, this is a IIA string theory
with coupling constant
$g_{IIA}=(R_{11}/l_p)^{3/2}$
compactified on a circle of radius $R_{10}$.
If both radii are small, then  the theory is IIB string theory with coupling
constant
$g_{IIB}=R_{11}/R_{10}$ compactified on a circle of radius
$R_{IIB}=l_p^3/R_{10}R_{11}$.
The limit in which $R_{10},R_{11}\to 0$ gives the IIB string in 10 dimensions
\cite{Asp}.
The moduli space is depicted in figure \ref{fig:duality1},
and should be identified under the reflection that interchanges $R_{10}$ with
$R_{11}$.

\begin{figure}[h]
\begin{center}
\scalebox{.5}
{
    \input{duality1.pstex_t}
}
\end{center}
\caption
{M theory compactified on a $T^2$ and the various string theory limits.}
\label{fig:duality1}
\end{figure}

The conjectured $\Z_2$ symmetry of M-theory used in the Ho\v rava-Witten
construction is
$I_{10}\Omega_M$
where $I_{10}$  takes $x^{10} \to - x^{10}$ and
$\Omega_M$ reverses the orientation of the M2-brane and the M5-brane and acts
in the supergravity
as $C \to -C$ where $C$ is the 3-form potential.
For M-theory compactified on a 2-torus with radii $R_{10},R_{11}$, this
symmetry reduces to
the various string theory symmetries considered in the last section in the
string theory limits.
If $R_{11} $ is large and $R_{10}$ is small, $I_{10}
\Omega_M$ acts as the symmetry $\hat \Omega$ of the IIA string, acting as
$(-1)^{F_L^c}$ in the
perturbative theory.
If $R_{10} $ is large and $R_{11}$ is small, on the other hand, $I_{10}
\Omega_M$ acts as the symmetry $I_{10}\Omega$ of the IIA string
compactified on a circle of radius $R_{10}$, where $\Omega$ is the IIA string
world-sheet parity
operator. If both radii are small and the IIB string is weakly coupled, so that
$g_{IIB}=R_{11}/R_{10}$ is small, $I_{10}\Omega_M$
acts as $\Omega$, the IIB string world-sheet parity
operator, while for strong coupling, the theory is the dual IIB string
theory with coupling
$\tilde g_{IIB}=R_{10}/R_{11}$ and
$I_{10}\Omega_M$
 is the IIB string symmetry $\tilde \Omega$, which acts as $(-1)^{F_L}$ in the
perturbative theory.
This is depicted in figure \ref{fig:duality2}.

\begin{figure}[h]
\begin{center}
\scalebox{.5}
{
   \input{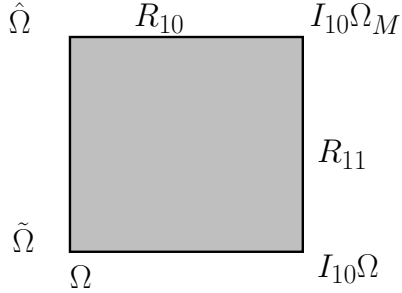}
}
\end{center}
\caption
{The Ho\v rava-Witten square, with the limiting
forms of the Ho\v rava-Witten symmetry arising in the various
string theory limits marked.}
\label{fig:duality2}
\end{figure}

The M-theory construction gives a non-perturbative picture of both the IIA and
IIB theories, and
shows that the symmetries discussed in the last section all extend to the same
symmetry of the
non-perturbative theory, namely the  Ho\v rava-Witten symmetry.
In particular, we see that $\tilde \Omega$ is
indeed the strong coupling limit of $\Omega$.

Consider M-theory on $T^2$ modded out by
$I_{10}\Omega_M$.
The circle in the $x^{10}$ direction becomes the interval $S^1/\Z_2$ and  the
torus is replaced by
a cylinder. It was argued in \cite{HW} that in the limit $R_{11} \to \infty$
and $R_{10}\to 0$ this
gives the $E_8\times E_8$ heterotic string,
in the limit $R_{10} \to \infty$  and $R_{11}\to 0$ this gives the type ${\rm
I}^\prime$ string and in the limit
in which
  $R_{11} \to 0$  and $R_{10}\to 0$, this gives the
type I string with coupling $g_{I }=R_{11}/R_{10}$ if this is small, and the
$SO(32) $ heterotic string with coupling
$\tilde g_{het}=R_{10}/R_{11}$ if this is small.
These are depicted in figure \ref{fig:duality3}.

\begin{figure}[h]
\begin{center}
\scalebox{.5}
{
   \input{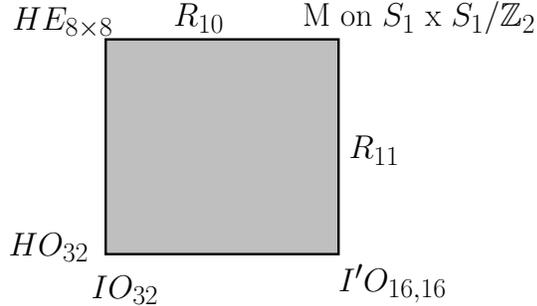}
}
\end{center}
\caption
{M-theory on a $T^2$ modded out with $I_{10}\Omega_M$ and the various string
theory limits.}
\label{fig:duality3}
\end{figure}

Comparing with the above, the Ho\v rava-Witten construction reduces to the
string theory
constructions of the last section in each of the string theory corners of the
moduli space.
In the corner in which $R_{10}$ is large and $R_{11}$ is small, the
Ho\v rava-Witten construction reduces to orientifolding the
IIA string with $I_{10}\Omega$ to obtain the type ${\rm I}^\prime$ string,
while in the corner
in which $R_{10}$ is large and $R_{11}$ is small, the
Ho\v rava-Witten construction reduces to modding out the
IIA string by $\hat \Omega \sim (-1)^{F_L^c}$ to obtain the
heterotic string.
If both  $R_{10}$   and $R_{11}$ are small, then
if $ R_{11}/R_{10}$  is small the construction gives the orientifold of the IIB
string with
$\Omega$ to get the type I string, while
if it is large it gives the IIB string modded out by
$\tilde \Omega \sim (-1)^{F_L }$ to give the SO(32) heterotic string.
However, in each of these constructions the 9-branes play a vital role and we
will now examine
the individual cases in more detail.

\subsection{$R_{10}$   and $R_{11}$ both   small}

For M-theory on $T^2$, the  limit in which both
radii tend to zero gives the IIB string with coupling constant given by the
limiting value of $g_{IIB}=R_{11}/R_{10}$ \cite{Asp}.
Modding out by $\Omega _M I_{10}$
gives    M-theory on a cylinder and we shall consider   the limits in which
both $R_{10}$ and
$R_{11}$ tend to zero; if $ R_{11}/R_{10}$ is small, this gives the type I
string with coupling
$g_I=R_{11}/R_{10}$, while if it is large it gives the $SO(32)$ heterotic
string with
$g_h=R_{10}/R_{11}$ \cite{HW}.

For M-theory on $T^2$, the symmetry $\Omega _M I_{10}$ reduces to  perturbative
symmetries
of the IIA or IIB theories in the corners of the    moduli space
square in Fig.~\ref{fig:duality1}.
  In particular, taking the limit in which $R_{10} \to 0,R_{11}\to 0$ with
 $ R_{11}/R_{10}$ small, $\Omega _M I_{10}$ becomes the symmetry $\Omega$ of
the IIB theory with
coupling  constant $g_{IIB}=R_{11}/R_{10}$ that is the fundamental string
world-sheet parity
symmetry, while taking the limit in which $R_{10} \to 0,R_{11}\to 0$ with
 $ R_{11}/R_{10}$ large, $\Omega _M I_{10}$ becomes the symmetry $\ti \Omega$
of the dual IIB theory
with coupling  constant $\ti g_{IIB}=R_{10}/R_{11}$ which acts in the
perturbative theory as
$(-1)^{F_L}$. Thus if $\Omega _M I_{10}$ is a symmetry of M-theory, then the
IIB string theory has a
non-perturbative symmetry which is the limiting form of $\Omega _M I_{10}$ in
the IIB limit, and
which acts as $\Omega$ at weak IIB coupling and as $\ti \Omega$ in the S-dual
IIB string.
Then M-theory modded out by $\Omega _M I_{10}$ becomes, in the limit, the
(non-perturbative) IIB
theory modded out by the $\ti \Omega / \Omega$ symmetry to give the SO(32)
heterotic/type I theory.
If $ R_{11}/R_{10}$ is small, this is the orientifold construction of the type
I theory and if it
is large, this is the construction of the SO(32) heterotic string of
\cite{Hull,Hnew}.

\subsection{$R_{10}$  large, $R_{11}$   small}

This limit gives the type ${\rm I}^\prime$ theory, with gauge group determined
by the positions of the D8-branes,
as discussed in section 5.3, and coupling constant given by (\ref{qwerty}).
For general D8-brane configurations,
the bulk supergravity theory between the
D8--branes is given by the
{\it massive} IIA supergravity, and  the limit $R_{10}\rightarrow \infty$
cannot be
taken, as the dilaton becomes infinite at some value(s) of $x^{10}$ as
$R_{10}$ is increased to a
finite critical value \cite{PW}.

However
in the special case in which 8 D8--branes positioned on top
of one orientifold
and 8 on top of the other, corresponding to the SO(16) $\times$ SO(16)
vacuum, there is no dilaton gradient and the limit
 $R_{10}\rightarrow \infty$ can  be
taken \cite{PW}.
In this limit, bound states of D0-branes with the O8-planes
become massless and lead to an enhancement of the gauge symmetry to $E_8\times
E_8$
\cite{Lowe,Sil,BGS}. This can be seen by first separating one D8-brane a
critical distance from each
O8-plane to get the type ${\rm I}^\prime$ theory with $E_8\times E_8\times
U(1)\times U(1)$ symmetry \cite{BGS}
and then taking the limit in which $R_{10}\rightarrow \infty$, in which the
critical separation
tends to zero so that all D8-branes coincide with the O8-planes in the limit.
At the orientifold planes, each D8-brane contributes one unit to the
cosmological  constant and the
total contribution is cancelled by the contribution of the O8--plane. The bulk
theory in the region
between the O8--planes is given by the {\it massless} IIA superstring,   which
has an
M-theory origin and so the decompactification limit can be understood
conventionally  in terms of
M-theory.

For finite $R_{10}$, we know that the theory has 16 moduli corresponding to the
positions of the
D8-branes, at least in the weak-coupling limit $R_{11}\to 0$, and there should
be an
interpretation for them at finite $R_{11}$. If there is such an interpretation,
it should
correspond to the positions (at particular values of $x^{10}$) of the M-theory
branes that give
rise to the D8-branes; these are the
M9-branes
 \cite{Hull,M9}, wrapped around the  $x^{11}$ circle. However, the bulk theory
between the D8-branes
is the massive string theory and so the M9-branes should arise in the M-theory
origin of the massive
string theory.
The  massive deformation of 11-dimensional
supergravity   proposed in
  \cite{BLO}
involves an explicit Killing vector (which is here $\partial
/\partial x^{11}$), and the IIA  configurations considered here can be lifted
to a solution of this theory consisting of
  a system of 16 M9-branes at various positions between two planes at the end
points of the
interval, and the positions can be arbitrary provided $R_{10},R_{11}$ are
finite.
We will refer to the fixed point planes that arise from oxidizing the
O8--planes as M-theory O9--planes.
The fundamental strings ending on D8-branes are oxidized to M2-branes ending on
the M9-branes.

\subsection{$R_{11}$  large, $R_{10}$   small}

This is the   construction of the heterotic string, compactified on a circle of
radius $R_{11}$,
from the IIA string (on a circle) with 32 NS-9A branes by modding out by $\hat
\Omega \sim
(-1)^{F_L^c}$. The gauge group is determined by the
heterotic string Wilson lines, which arise here as the
16 moduli given by the expectation values of
the scalars
$c^{(0)}$ in a Cartan subalgebra of SO(32).
This is related by T-duality in the $x^{11}$ direction to the SO(32) heterotic
string
constructed from the IIB string, where the gauge sector arose from open
D-strings with one end
on the
fundamental heterotic string and the other on a D9-brane, which collapse to
zero length at weak
coupling. After the T-duality, the open D-strings become D0-branes and open
D2-branes ending on the
NS-9A branes. The D2-branes are cylindrical, wrapping the circular $x^{11}$
direction and with one
end on the fundamental string in the 9-dimensional space orthogonal to $x^{11}$
and the other on an NS-9A brane, and the length of the cylinder tends to zero
at weak coupling.
It is these D0 and D2 branes that are responsible for the gauge sector at weak
coupling.

At finite heterotic coupling, the non-perturbative
theory is M-theory on a cylinder, with an eleventh dimension which is a finite
interval of length
$R_{10}\pi$.
The NS-9A branes oxidize to 16 M9-branes, moving between two O9-planes
with the 16 moduli now giving    the positions of the M9-branes on the interval
(or to 16 M9-branes
plus their mirror images on the circle). The non-BPS D2-branes ending on NS-9A
branes become
cylindrical M2-branes ending on M9-branes  and orthogonal to the $x^{10}$
direction
and the
fundamental IIA strings become M2-branes that are tangential to the $x^{10}$
direction, and again
end on M9-branes.
The  heterotic string in 10 dimensions results from the decompactification
limit $R_{11}\to \infty$ with all the M9-branes at one O9-plane, in which case
the SO(32) heterotic string arises,  or with 8 M9-branes at each of the
O9-planes, giving the $E_8 \times E_8$ heterotic string.

In the limit $R_{11}\to \infty$  with 8 M9-branes
at each O9-plane, the M9-branes can no longer move. This is in accord with the
fact that the tension
of a single M9--brane is proportional to $(R_{11})^3$  (see Figure 2) so that
they become
infinitely massive in the limit, and that their world-volume theory has no
scalar and so no
collective coordinate for translations in the transverse direction.

\subsection{M9-Branes and the Ho\v rava--Witten Construction.}

M-theory compactified on a line interval in the $x^{10}$ direction of length
$R_{10}\pi$
gives a non-perturbative formulation of the $E_8\times E_8$ heterotic string
and
requires an $E_8$ super-Yang-Mills theory on each end-of-the-world
plane  \cite{HW}. If this is further compactified on a circle in the
$x^{11}$ direction, then we have learned that each end-of-the-world
plane consists of an O9-plane and 8 M9-branes which can move away from the
O9-plane if $R_{11}$ is
finite.
This is because we know that as $R_{11} \to 0$ this gives the IIA theory on an
interval,
and the two end-of-the-world
planes are each O8-planes with 8 D8-branes that can move away from the
O8-plane, and for finite
$R_{11}$ this must oxidise to  M9-branes wrapped on the $x^{11}$ circle moving
between O9-planes.

The world-volume theory of the 8 M9-branes at an O9-plane has the $E_8$
super-Yang-Mills theory
on $R^{8,1}\times S^1$
 as
its low-energy limit and moving
  the  M9-branes in the $x^{10}$ direction breaks
the gauge group, as the M9-brane positions  are moduli that correspond to
Wilson lines in the $x^{11}$ direction.
The compact $x^{11}$ direction plays a special role in the theory, which has
explicit dependence on
the Killing vector $k=\partial /\partial x^{11}$. The theory between the
M9-branes is the massive
version of M-theory depending explicitly on $k$ \cite{BLO}, and the M9-brane
world-volume effective
action also depends explicitly on $k$. Dependence on $k$ and on the mass
parameter drops out if 8
M9-branes are placed at each O9-plane, and it is only in this configuation that
the
limit $R_{11}\to \infty$ can be taken, to give the Ho\v rava--Witten
configuration.

\bigskip

\section{Conclusions}

In this paper we  have discussed the spacetime--filling branes of Type IIA and
Type IIB string theories, which
are    connected by a web of
duality transformations and/or reductions
(see Figure \ref{fig:branes}).
In the first part of this work
we have constructed their worldvolume effective actions,
together with their reductions to nine dimensions.

In the second part of this work we  investigated the roles of D9, NS-9A and
NS-9B branes in string theory.
Each of these lead to inconsistencies when introduced in a type II string
theory,
but in each case 16 of these branes plus 16 mirrors can be added to
the type II
string theory modded out by the appropriate $\Z_2$
symmetry to obtain a construction of string theories with 16 supercharges:
the IIB theory modded out by $\Omega$ with 32
D9-branes gives the type I string, the IIA string modded out by $I_{10}\Omega$
with 16(+16) D8 branes gives the type ${\rm
I}^\prime$ string,  the IIB string modded out by $\tilde \Omega$ with 32 NS-9B
branes gives the SO(32) heterotic string and
the
IIA string compactified on a circle and modded out by $\hat \Omega $ with 32
NS-9A branes gives the heterotic string
compactified on a circle. In the latter case, the large radius limit can be
taken only in the $O(16)\times O(16)$ vacuum, to
obtain the
$E_8\times E_8$ heterotic string from the IIA string.
These constructions are related to one another by dualities
(see Figures \ref{fig:strings} and \ref{fig:z2}).

Moreover, these four constructions each arise as a different limit of the Ho\v
rava-Witten construction, and so are unified
in M-theory. This gives the origin of the dualities relating these
constructions.
These connections also show that, when compactified on a circle of radius $R$,
the
  end-of-the-world-branes of Ho\v rava and Witten
each become a combination of an O9-plane and 8 M9-branes, all wrapped on the
circle, and the
8 M9-branes can be moved to arbitrary points on the interval $S^1/\Z_2$.
In the small radius limit $R\to 0$, these become the D8-branes and O8-planes of
the type ${\rm
I}^\prime$ string theory.
However,   the large radius limit $R \to \infty$ does not exist unless 8
M9-branes are at each O9-plane, in which case the
Ho\v rava-Witten picture is recovered.

Our work implies that the group structure of the $E_8 \times E_8$
string theory can be described by D0-branes and open non-BPS D2-branes which in
the
strong coupling limit oxidize to open M2-branes ending on the
M9--branes. These are the analogues of the  D1--branes that
provide the group structure in the case of the SO(32) Heterotic string theory.

Finally, we have found that if the orbifolding of the IIB (IIA) string by
$(-1)^{F_L}$ to obtain the IIA (IIB) string
were to extend  to the full non-perturbative theory, it would have some
peculiar features; for example, the orientifolding of
the IIB string {\it without} introducing D9-branes would lead to M-theory. It
would be of considerable interest to
investigate  whether or not this construction does extend to the
non-perturbative theory, and if so, how its peculiar
features can be understood directly.


\section*{Acknowledgments}

We would like to thank M.~Green, L.~Huiszoon,
D.~Matalliotakis, N.~Obers,
B.~Pioline and M.~de Roo for useful discussions.
E.B.~likes to thank A.~Karch for a
useful discussion about the relation with the Ho\v rava-Witten picture.
E.B., C.M.H. and Y.L.
 would like to thank the organizers and participants of the
1998 Amsterdam Summer Workshop on {\it String Theory and Black Holes}
for providing a stimulating environment.
The work of E.B.~is supported by the European Commission TMR program
ERBFMRX-CT96-0045, in which E.B.~is associated to the University of
Utrecht. The work of E.E, R.H.~ and J.P.v.d.S.~ is
part of the research program of the ``Stichting voor Fundamenteel
Onderzoek der Materie'' (FOM).
The work of C.M.H. ~is supported by an EPSRC senior fellowship.

%
%

\end{document}